\documentclass[11pt]{article}
\usepackage{graphicx}
\usepackage{subcaption}
\usepackage{color,soul}
\usepackage{amsmath,amssymb}
\usepackage{authblk}
\usepackage{float}
\usepackage{tikz}
\usepackage{tikz-network}
\usepackage{tikzpeople}

\title{Localized risk perception triggers early behavioral adaptations in epidemics on networks}

\author[1]{Baltazar Espinoza}
\author[2]{Jimmy Calvo-Monge}
\author[2,3]{Fabio Sanchez}
\author[4]{Simon A. Levin}
\author[1]{Madhav Marathe}

\affil[1]{Biocomplexity Institute, University of Virginia, Virginia, USA}
\affil[2]{CIMPA, Universidad de Costa Rica, Ciudad Universitaria Rodrigo Facio, San José, 11501, Costa Rica}
\affil[3]{Escuela de Matemática-CIMPA, Universidad de Costa Rica, Ciudad Universitaria Rodrigo Facio, San José, 11501, Costa Rica}
\affil[d]{Department of Ecology and Evolutionary Biology, Princeton University, Princeton, NJ 08544, USA.}

\date{\today}

\begin{document}
\maketitle

\begin{abstract}
The contact structure of the population shapes the progression of epidemics. Nonetheless, the joint evolution of individual behavioral adaptations and disease dynamics on networks remains poorly understood. We use a behavioral-epidemiological model to study the joint evolution of human behavior and epidemic dynamics on networks. Our results reveal how the adaptation of local social structures, influenced by risk–benefit trade-offs, affects the dynamics of epidemics. We allow the epidemic and population-level behavior dynamics to emerge from the heterogeneous behavioral responses of individuals. Our framework assumes that individuals adjust their contact structure by temporarily dropping or maintaining connections based on perceived benefits and risks. Our results show that behavioral responses induced by localized risk perceptions lead to premature population-level responses relative to epidemic dynamics. Specifically, individual efforts peak at the epidemic maximum, while population-level efforts remain modest. We explore the robustness and extensions incorporating heterogeneous subpopulations.
\end{abstract}

\section*{Keywords}
Epidemiology; adaptive behavior; epidemic network models; individual behavior

\section{Introduction}

Adaptive processes in complex systems are known to yield population-scale dynamics emerging from individuals' interactions~\cite{Simon03}. In the case of epidemics, accurately modeling the intertwined dynamics between global dynamics and individual processes presents a significant and complex challenge. For example, mean-field models are mathematically tractable and provide valuable epidemiological insights, with the potential to include a certain level of heterogeneities. However, these models often overlook processes modulating transmission across scales. The study of individual incentives and decisions, and their subsequent impact on epidemic dynamics, requires complex modeling mechanisms~\cite{Traulsen23}.
Modern mathematical models incorporating network structures have the capability of addressing many of the complexities driving transmission processes.
Network models enable the study of the feedback between individuals' attributes, the underlying structure of social interactions, over which the transmission process evolves, and the epidemic dynamics~\cite{Danon2011,Keeling2005,Kiss17}.
%
Usually, these models are classified as evolving and adaptive networks~\cite{Guo13}. Evolving networks focus on the impact that changes in the network topology have on the transmission dynamics. In these models, the underlying network's topology is crucial to studying the disease transmission process.
In the seminal paper~\cite{albert2000error}, Albert {\it et al.} showed that the intrinsic network topology has profound consequences on error tolerance and survivability, which has critical implications on embedded transmission processes. Likewise, Gates {\it et al.}~\cite{GATES201511}, showed that it is possible to induce favorable disease transmission behavior by manipulating the underlying topological nature of the contact networks.
In these studies, changes in the network topology are phenomenologically incorporated and linked to the progression of the epidemic. In other words, the impact of the network's topology change is modeled as an exogenous factor influencing the progression of the epidemic\cite{leventhal2015evolution,rocha2013bursts,vazquez2016rescue}.

On the other hand, in adaptive networks, the network's topology is changed in response to the disease dynamics. Different adaptive approaches and methodologies to epidemic network modeling have been studied~\cite{Bansal10,Pastor15}. These modeling frameworks exhibit complex transmission dynamics and use network topology alteration processes. For example, Gross {\it et al.}~\cite{gross2006epidemic} showed that changes on the network's topology produced by a probabilistic-based rewiring process (dropping edges connecting susceptible to infectious nodes and creating new susceptible-to-susceptible edges), make complex transmission dynamics. Moreover, Shaw {\it et al.}~\cite{shaw2010enhanced} showed that the interaction between the rewiring process and vaccination facilitates disease control in adaptive networks compared to static networks.
The processes of rewiring, link switching, or random link activation and deletion are the primary strategies to affect the network structure in most of these frameworks~\cite{Ball2022,Guo13,Kiss2012,Ogura2017,RISAUGUSMAN200952,Zhu2019}. Nonetheless, in some cases, rewiring process strategies are not sufficient to control epidemics~\cite{DONG2015169}. Rewiring mechanisms can be governed by epidemic dynamics or network components; for example, Shivakumar {\it et al.} \cite{Shivakumar12}, used a preferred degree approach, in which nodes alter their contacts to fit an expected network degree.

These modern adaptive network models employ phenomenological approaches to incorporate network changes as a surrogate for studying the impact of human behavioral responses during epidemics. 
Such models disregard the relationship between the individual-scale behavioral incentives and mechanisms, and the observed large-scale dynamics~\cite{Bansal07,Kao2010}.

Furthermore, in many instances, the introduction of complex rewiring methods can be unrealistic or non-optimal, depending on the individuals' attributes. For example, reducing contacts might be more advantageous for the agent than substituting them, or a delay may be required to successfully perform a rewiring process~\cite{Li19}. 
Another critical aspect usually overlooked in these studies 
is the impact of individuals' contact reduction. In reality, individuals need to engage in contacts to obtain benefits; therefore, the process of changing the network topology should take into account the individuals' contact requirements. For instance, the spatial network model proposed in~\cite{Maharaj12} includes socio-economic considerations in the contact reduction process. 


In this study, we propose a novel mechanistic methodology that incorporates the feedback between epidemic progression and individuals' decentralized adaptive behavioral responses within an adaptive network setting.
We consider the population-scale behavioral and epidemiological outcomes as observable consequences of the interaction between epidemic progression and heterogeneous behavioral responses driven by individuals' local perceptions of the infection risk.
This modeling framework explicitly incorporates individual-level adaptive behavioral processes by modeling changes in the network topology as a result of the adapting nature of the underlying decision-making processes. 

\begin{figure}[!h]
    \centering
    \includegraphics[width=\linewidth]{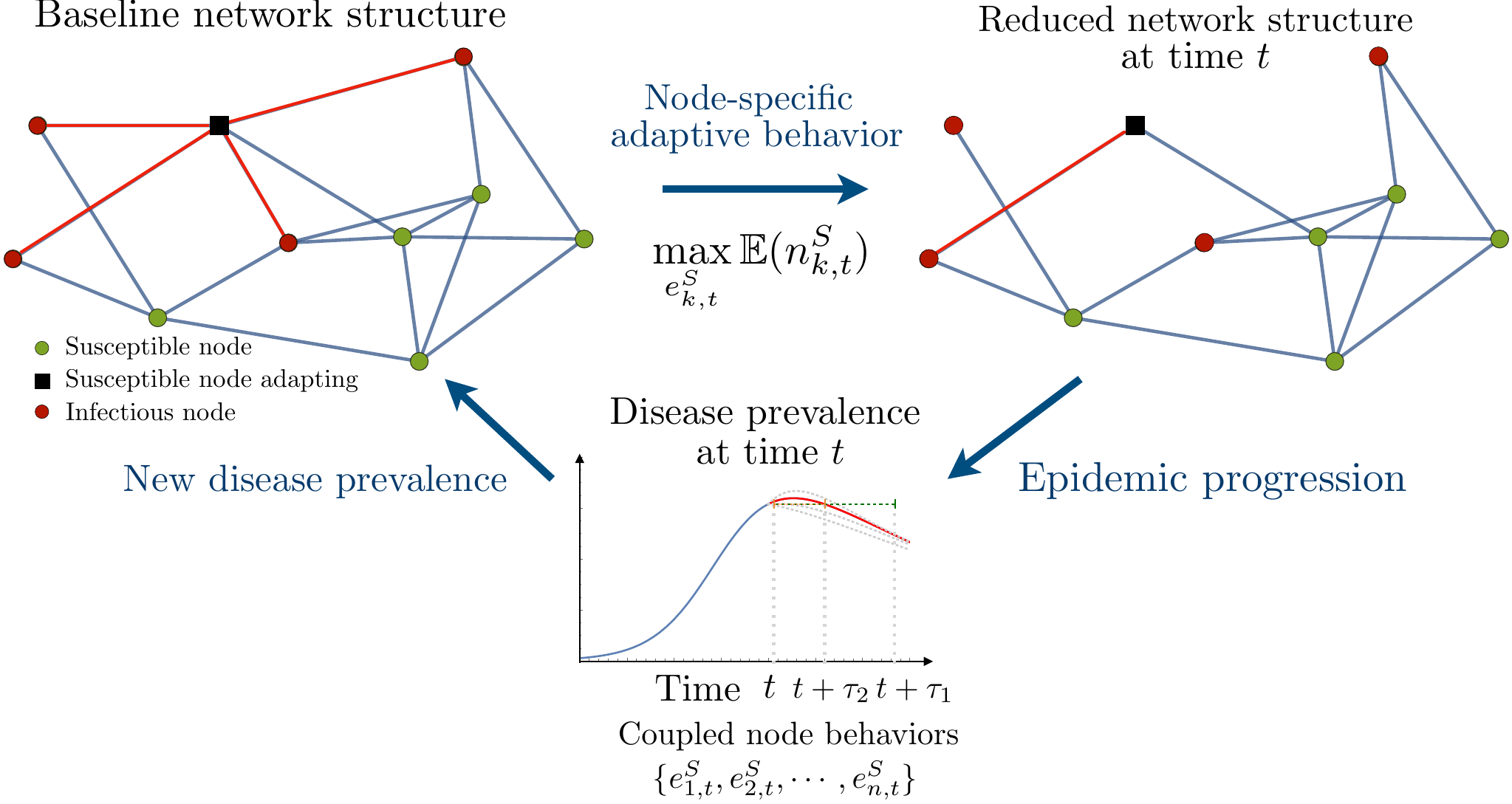}
    \caption{
    {\bf Schematic of the modeling framework components.}
    We couple an epidemic model and a model of heterogeneous adaptive human behavior driven by local network heterogeneities.
    Individuals adjust their number of edges based on the local infection risk they perceive, which is determined by the number of infected neighbors. We assume that individuals are unaware of the decisions made by others and are unaware of the infection risk they pose to others.
    The aggregation of adaptations constitutes the population-level behavioral response that reshapes the epidemic trajectory.
    }
    \label{fig:framework}
\end{figure}

As local interaction conditions continuously change with the progression of the epidemic, individuals choose their privately optimal behavior based on the balance between the benefits of maintaining social interactions and the associated risk of infection.
%
%
By incorporating spatial resolution of network structures, we extend the existing framework of adaptive human behavior in epidemics originally formulated for mean-field models of disease transmission~\cite{espinoza2021adaptive,espinoza2024adaptive,espinoza2022heterogeneous,fenichel2011adaptive,morin2013sir,perrings2014merging} to study how aggregated behavioral and disease dynamics emerge from individuals' asynchronous behavioral responses as the epidemic propagates through the population.
The main technical innovations of the proposed modeling framework span two directions:
($i$) The forward-looking formulation of the decision framework incorporates individuals' expectations of future benefits, future infection risk, and potential future transitions to alternative health states;
($ii$) The coupling of node-specific {\em Markov decision processes} (MDPs) to model heterogeneous adaptive behavioral responses.
In the proposed model of adaptive behavioral responses during epidemics on networks, susceptible nodes optimize their social interactions by changing their neighborhood structure based on their private benefits and risk perceptions. 
The node-specific optimal neighborhood structure is determined by temporarily adjusting the number of edges of susceptible nodes.
Our model framework is schematically outlined in Figure~\ref{fig:framework}, and the detailed formulation of the behavioral model through the local network structure optimization problems is described in the Methods section.

Our approach differs from the previously discussed frameworks in that we do not consider the creation of new edges. Instead, the dynamics of our adaptive network rely on temporarily dropping and maintaining edges, based on the original network topology. The network structure is adjusted at each time step as a result of the local optimization processes of the susceptible nodes.
Moreover, while diverse approaches utilize the current and past states of the system to guide behavioral decisions \cite{epstein2008coupled,funk2009spread,reluga2010game}, our proposed framework envisions individuals' decisions as dependent on both the current system's state and a projection of the future.
The node-specific decision process incorporates both the expected future benefits and the risk of infection, which depend on the current infection risk and the potential future transitions to alternative health states. The influence of future expectations on the node-specific decision process is modeled by projecting the system's state over a finite planning horizon.
The network-based formulation of the behavior-epidemic model enables us to explore both individual-level and population-scale behavioral responses. 
Additionally, we show that the proposed framework can incorporate heterogeneous behavioral profiles and study the aggregated impact of these during epidemics.

\section{The interplay between epidemics and adaptive behavior on networks in homogeneous populations}
Since the proposed model is not analytically tractable, we investigate the impact of individual-level behavioral responses on both the epidemic dynamics and the adaptive network structure through numerical experiments.
We track the {\em population-level behavioral response}, which emerges from the aggregation of individual-level behavioral choices, by computing the total number of edges on the dynamic network at each time step. Additionally, we track the {\em individual-level behavioral response} by computing the average number of edges maintained by susceptible nodes at each time step. The detailed formulation of the behavioral response framework is provided in the methods section.

\subsubsection{Individual- and population-level adaptive behavioral responses occur asynchronously during epidemics}
\begin{figure}[ht!]
\centering
\includegraphics[scale=0.4]{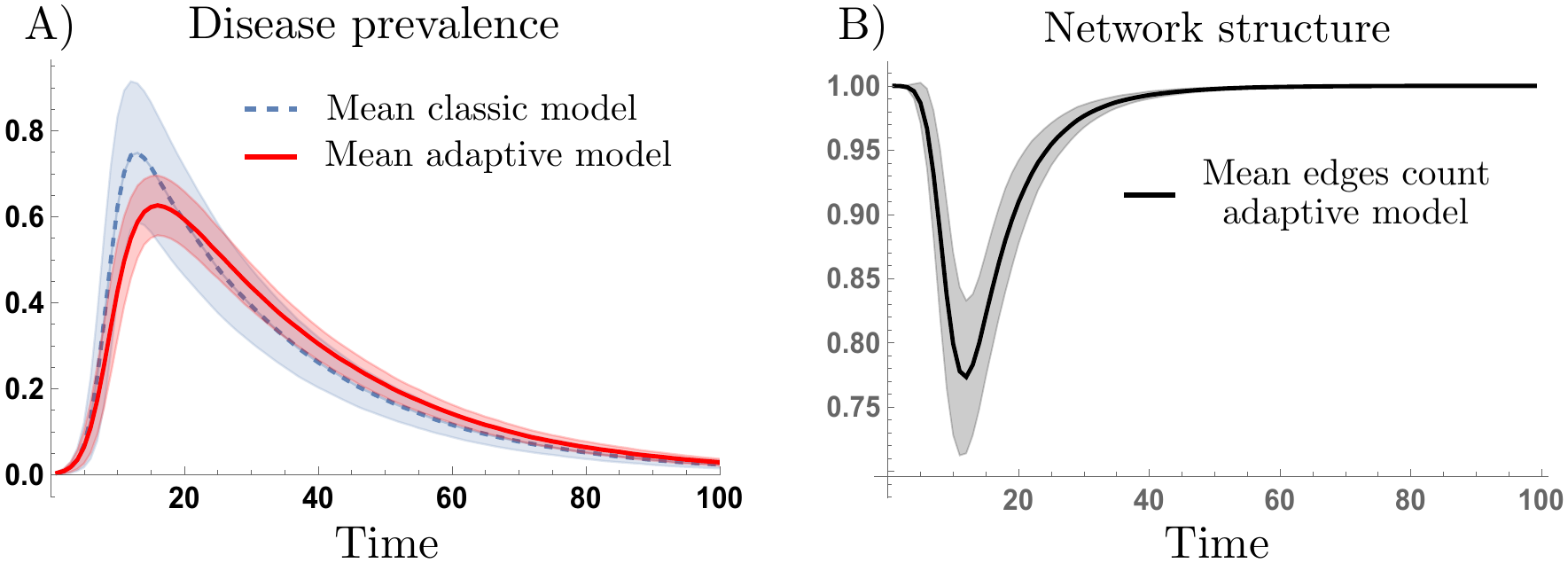}
\caption{
    {\bf Adaptive behavioral responses decrease the epidemic peak and dynamically change the network structure.}
    (Panel~A) Disease dynamics for the static network model (dashed line) and the adaptive behavior model (thick line). The adaptive responses of susceptible nodes, which modulate the network's structure, reduce and delay the epidemic peak.
    (Panel~B) The network's structure during the progression of the epidemic. The normalized mean number of edges in the social network decreases around the time of the epidemic peak and then increases back to its original structure as the epidemic subsides.
    We assume a population's risk acceptance of $\nu=0.05$ and a planning horizon $\tau=7$ days.
    }
 \label{fig:tseries-netdyns}
\end{figure}

We study the intertwined dynamics of epidemics and behavior, assuming that susceptible nodes exhibit homogeneous sensitivity to the perceived risk of infection. Our focus is on the dynamics of individual- and population-level behavioral efforts as a metric for studying the impact of the contagion process across scales. Figure~\ref{fig:tseries-netdyns} shows the mean and the interquartile boundaries of the disease dynamics for both the classic and the adaptive model, along with the corresponding adaptive network dynamics. Figure~\ref{fig:tseries-netdyns}A shows that the adaptive model produces a delayed and smaller prevalence peak (thick line), relative to the epidemic dynamics exhibited in the static network model (dashed line). Figure~\ref{fig:tseries-netdyns}B illustrates how the adaptive network structure gradually reduces and rebounds in the mean number of edges as the epidemic progresses.

\begin{figure}[ht!]
\centering
\includegraphics[scale=0.4]{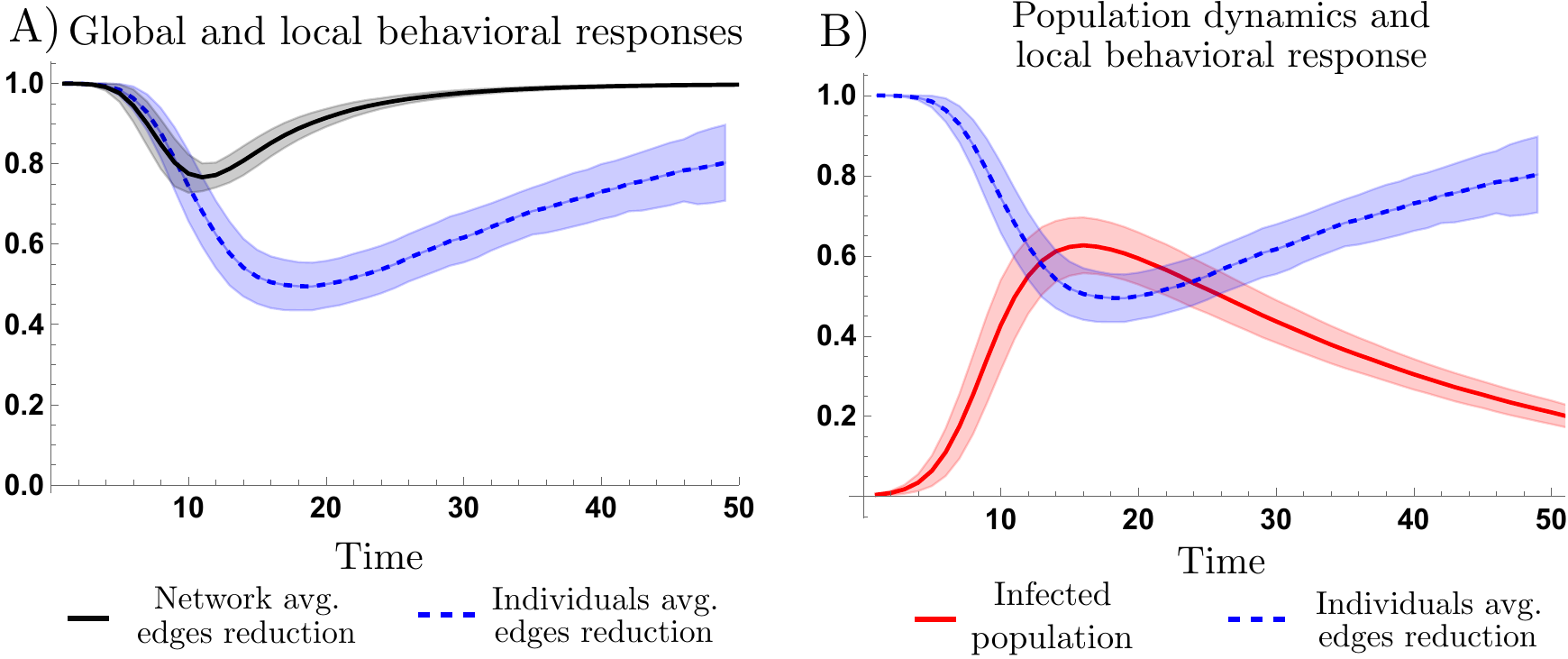}
\caption{
    {\bf The individual- and the population-level behavioral responses asynchronously occur as the epidemic propagates on the network.}
    (Panel~A) Individual-level (dashed blue curve) and population-level (thick black curve) behavioral responses.
    The population-level behavioral response occurs earlier than the individual-level one. Moreover, the perceived population-level behavioral effort (reduction of nodes) is lower than the individual-level efforts.
    (Panel~B) Disease prevalence (red thick curve) and individual-level behavioral response (dashed blue curve). The mean individual-level behavioral effort reaches its maximum (or minimum number of edges in the adaptive network) at the time the epidemic peak occurs.
    We assume a population's risk acceptance of $\nu=0.05$ and a planning horizon $\tau=7$ days.
    }
 \label{fig:tseries-edgsred}
\end{figure}

The significant result here is that the pace and intensity of behavioral responses differ significantly across scales (Figure~\ref{fig:tseries-edgsred}A). 
Specifically, the maximum behavioral effort at the population level occurs before the epidemic peak, asynchronously with the maximal individual-level behavioral efforts, which attain their maximum simultaneously with the epidemic peak (Figure~\ref{fig:tseries-edgsred}B).

This represents a key difference from the previous results obtained in the analogous mean-field adaptive behavior framework, where the population-level maximal behavioral responses coincide with the epidemic peak~\cite{espinoza2021adaptive,espinoza2022heterogeneous,fenichel2011adaptive,perrings2014merging}.
The reasoning behind the asynchronous behavioral dynamics across scales lies in the trade-off between the size of the population that modifies its behavior and the magnitude of the efforts required.
Consequently, the perceived population-level behavioral effort is maximal not during the epidemic peak, when a small proportion of the population is ardently trying to evade infection, but rather when many susceptible nodes make relatively small efforts to avoid infection.
We found that the asynchronous and distinct behavioral response dynamics across scales are driven by both the number of susceptible nodes adapting their behavior simultaneously and the asynchronous infection risk faced by susceptible nodes as the epidemic propagates over the network.
The observed phenomenon that the population-level optimal behavioral response is not attained at the critical time during the epidemic emerges as a combination of the volume of susceptibles in the stages before the epidemic peak and the stress placed on a diminishing number of susceptible individuals as the disease progresses.
Our results demonstrate that decentralized adaptive behavioral responses based exclusively on local information significantly impact the disease dynamics. However, the lack of coordinated efforts inherently limits the population's ability to ameliorate the epidemic burden without centralized efforts.
We show that when risk avoidance behavior is driven by local information, the individual-level optimal behavioral choice does not represent the best response at the population level. In these scenarios, the absence of global information leads individuals to overlook the global situation and consequently the population-level optimal behavioral decisions.

\subsubsection*{Individual incentives and the network structure modulate behavioral efforts across scales}
To study the robustness of our results, we explore the impact of behavioral variations over different network structures.
In the proposed model, the observed behavioral response difference across scales and the consequent impact on disease dynamics are directly influenced by the heterogeneous neighborhood structure.

On one side, our behavior model captures individual incentives via two main model parameters: 
$(i)$ the risk acceptance level 
($\nu$), and $(ii)$ and the planning horizon ($\tau$); on the other side, the network connectivity plays a principal role in determining the progression of epidemics.

\begin{figure}[!h]
  \centering
  \begin{subfigure}[b]{0.48\linewidth}
    \centering
    \includegraphics[width=\linewidth]{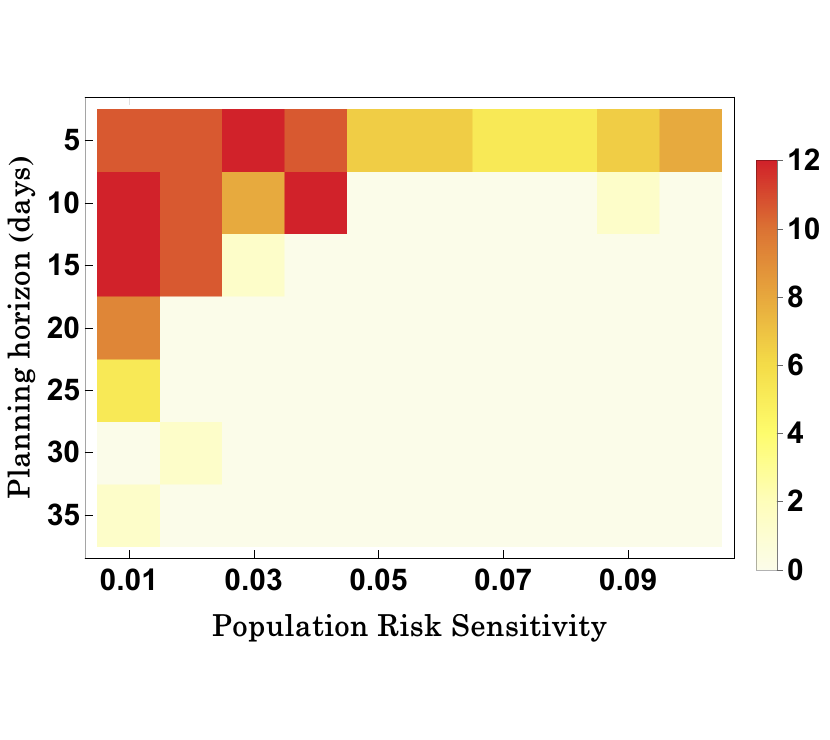}\  
    \caption{Effect of $(\nu,\tau)$ on global-local day lag.}
    \label{fig:heatmaps_A}
  \end{subfigure}
  \hfill
  \begin{subfigure}[b]{0.48\linewidth}
    \centering
    \includegraphics[width=\linewidth]{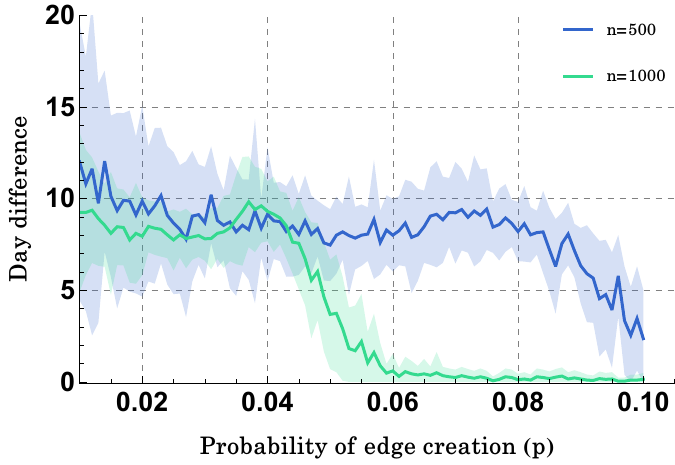}\  
    \caption{Effect of network connectivity on global-local day lag.}
    \label{fig:heatmaps_B}
  \end{subfigure}
  \caption{The average day difference between the maximum edge reductions at local and global levels. Simulations use an Erd\H{o}s–R\'{e}nyi graph with $N=500$ and edge probability $p=0.05$.}
  \label{fig:heatmaps_figure}
\end{figure}
Figure~\ref{fig:heatmaps_figure}A shows that by increasing individuals' infection risk acceptance or the planning horizon period, the effort made by susceptible individuals to avoid contacts reduces. In these scenarios, the maximum risk avoidance efforts at the global and local scales tend to synchronize.

Our numerical experiments show that the sensitivity to infection risk modulates the difference in contact efforts observed across scales in Figure~\ref{fig:tseries-edgsred}.

The topology of the underlying network is a crucial factor in determining disease behavior for network epidemic models \cite{Ganesh05,Meyers07,Perez22}. 

Specifically, the availability of contacts and how closely connected these contacts are to each other are defined by the networks' connectivity.

Figure~\ref{fig:heatmaps_figure}B shows the average lag between the maximum contact reductions at the individual and population scales, for different connectivity levels of the underlying Erd\"{o}s–R\'{e}nyi network of $n$ nodes.

Our results are robust to variations in the underlying network model. Particularly, in sparse networks, the adaptive behavior model reveals a significant discrepancy between behavioral responses at the individual and population scales.
We examine the sensitivity of our results using scale-free~\cite{Barabasi99} and small world~\cite{Strogatz98} underlying network structures in the SI Appendix.

\section*{The interplay between epidemics and adaptive behavior on networks in heterogeneous populations} 
\begin{figure}[ht!]
  \centering
  \begin{subfigure}[b]{0.48\linewidth}
    \centering
    \includegraphics[width=\linewidth]{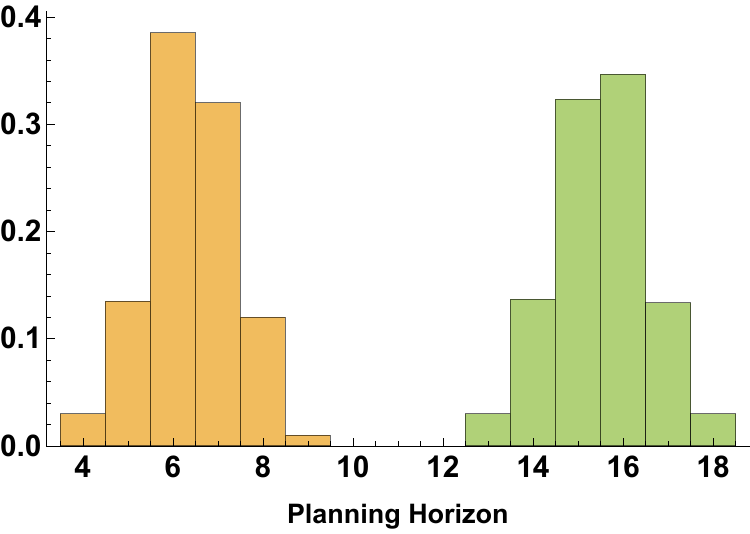}
    \caption{Population's planning horizon distribution}
    \label{fig:distplotT}
  \end{subfigure}
  \hfill
  \begin{subfigure}[b]{0.48\linewidth}
    \centering
    \includegraphics[width=\linewidth]{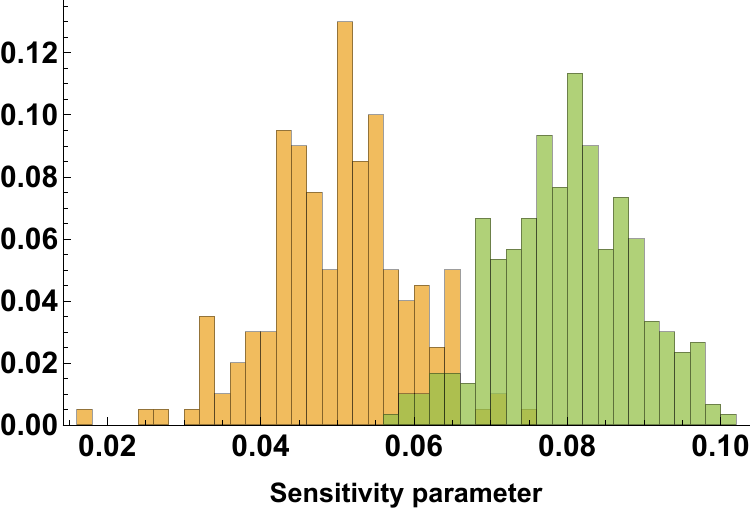}
    \caption{Population's risk sensitivity distribution}
    \label{fig:distplotV}
  \end{subfigure}
  \vspace{6pt}
  \begin{subfigure}[b]{0.48\linewidth}
    \centering
    \includegraphics[width=\linewidth]{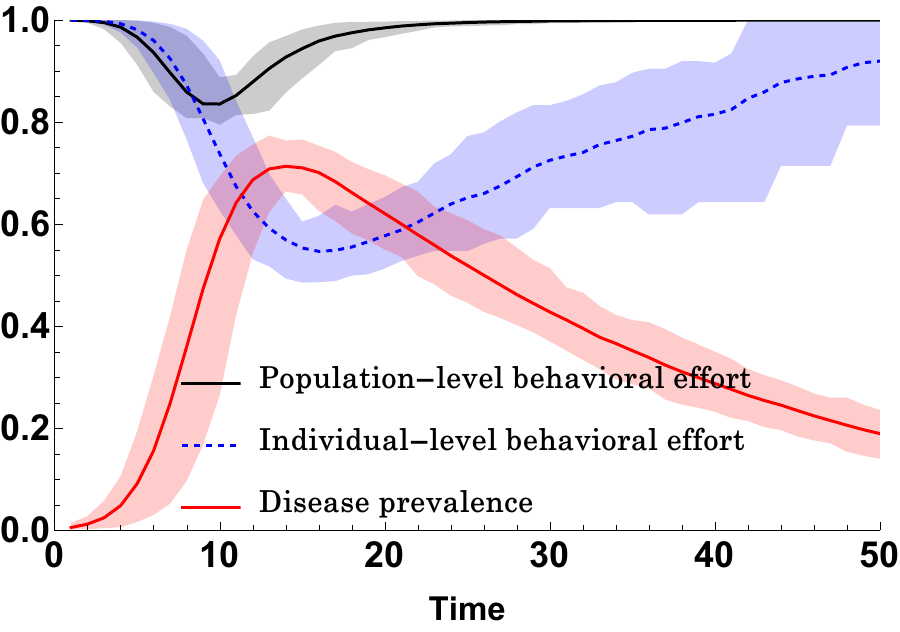}
    \caption{Global-Local behavior with population distributions}
    \label{fig:global_local_dist}
  \end{subfigure}
  \hfill
  \begin{subfigure}[b]{0.48\linewidth}
    \centering
    \includegraphics[width=\linewidth]{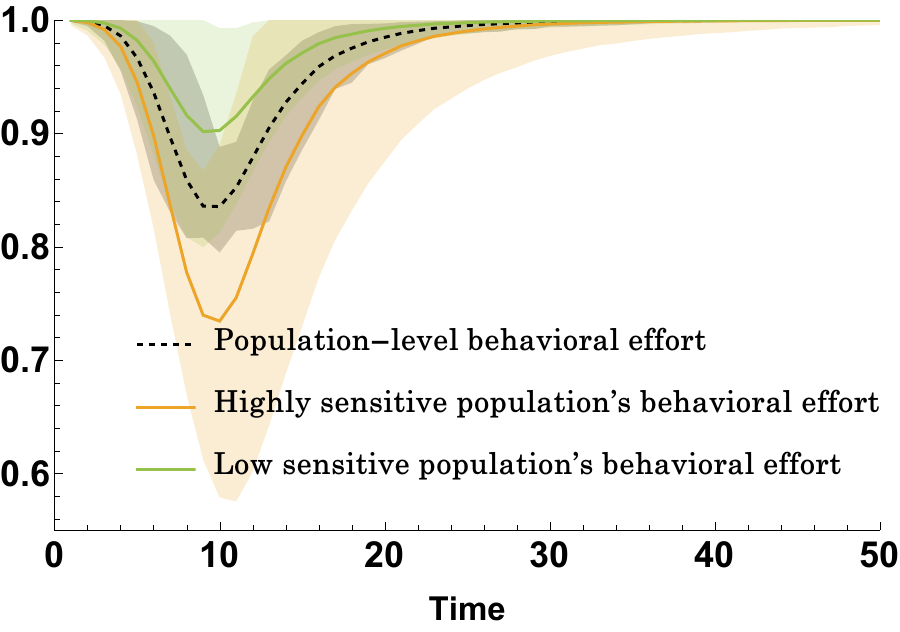}
    \caption{Contact difference between sub-populations}
    \label{fig:subpop_resp}
  \end{subfigure}
  \caption{On panels (a) and (b), distributions for adaptive parameters for two sub-populations. In yellow, sub-population 1 has a lower planning horizon and utility curvature (higher sensitivity). In green, sub-population 2 has higher parameters (lower sensitivity). Panel (c) shows that the global/local behavior effort lag persists under distributions of adaptive parameters. Panel (d) shows the difference in contact behavior for each sub-population.}
  \label{fig:distribution_figure}
\end{figure}

In this section, we explore the impact of adaptive behavior on populations composed of individuals with heterogeneous risk tolerance, driven, for instance, by distinct socio-economic conditions.
%
%
We incorporate variations in the population's characteristics, assuming bimodal distributions of planning horizons and risk sensitivities.
We assumed a population composed of two behavioral groups. For instance, a sub-population with a high poverty index, where few people would be economically prepared to plan for long periods, and a sub-population with a low poverty index that may afford the expenses of planning for long periods.
Figure~\ref{fig:distribution_figure}A and B show the normalized distributions of the planning horizon and risk sensitivities associated with the heterogeneous behavioral profiles in the population.
Figure~\ref{fig:distribution_figure}C shows that the behavior discrepancy across scales is robust to these populations' heterogeneities. 
Finally, Figure~\ref{fig:distribution_figure}D shows the differences in contact behavior for each population. Our results reveal that behavioral heterogeneity only impacts the stringency of the sub-populations' behavioral responses, ultimately modulating the population-level behavioral effort, the average of both sub-population responses.
Similar to our baseline scenario, our results show the same behavioral phenomenon across scales, where the population is assumed to show homogeneous behavioral profiles.
That is, the asynchronous behavioral efforts observed across scales are robust to variations in the population's structure.


The underlying network model can also provide us with a myriad of scenarios whose modeling process can be enhanced by incorporating the behavioral-epidemiological trade-off implemented by the adaptive framework. For example, the disaggregation of contact decision processes by social layers, where agents have different attitudes towards engaging with close and regular neighboring nodes and towards doing so with sporadic or stranger contacts, can be found in an example of a multi-layer contact network model~\cite{Network2}. This difference in local risk perceptions between contact layers offers an application of the main adaptive framework. We briefly explore these more complex contact decision dynamics in the supplementary information.

\section*{Discussion}


%
The containment of epidemics is intrinsically tied to our capacity to control localized outbreaks and modulate how these aggregate through already established patterns of mobility. However, in the last instance, disease containment depends upon individuals' behavioral decisions governing social interactions, whose aggregation drives the population-level contagion process. For this reason, it is critical to understand the epidemiological consequences of coupling epidemic network dynamics and decision processes driven by individual-level trade-offs.

The behavioral-epidemiological framework proposed in this study allows us to study the impact of individual-level decision dynamics on the aggregated epidemiological landscape. Specifically, we examine the effects of individual incentives (risk aversion and economic considerations) on the network's topology evolution process, and ultimately on the progression of epidemics.
This mechanistic framework avoids contact adaptation methods that are not intrinsically connected to disease progression; instead, it models the decision-making process using a novel approach in which socio-economic and epidemiological factors play a key role in adjusting the network structure during epidemics. Moreover, the proposed adaptive framework enables the consideration of heterogeneous populations with specific individual-level behavioral profiles, offering a powerful tool to better reflect the diverse attitudes towards the progression of epidemics among the affected population.

Beyond the proposition of the novel framework, our main result is the asynchronous nature between the behavioral stress observed at the population level and that observed at the individual level. We found that locally based adaptive behavior results in an early global maximum effort.
The key insight of our results is that the global maximum effort is not attained by the individual-level maximal behavioral efforts; instead, it is driven by a big population size performing medium-level risk avoidance efforts.
In contrast, during the peak of the epidemic, the global behavioral effort is carried out by the remaining few susceptible individuals, who face a high risk of infection.
Our modeling framework allows us to show that the global disease dynamics are modulated by the coupling between the epidemic progression over a dynamic underlying network (driven by behavioral responses and incorporated via edge regulation), and the individual-level behavioral dynamics (driven by disease progression and neighborhood structures, incorporated via risk perception and planning horizons).
%

Further numerical explorations in the SI appendix 
%
show that our results are robust to variations in the population's risk sensitivity and network structure, which may emerge due to factors such as socioeconomic characteristics, beliefs, education, and demographics.
Specifically, our baseline implementation utilizes the health status of immediate neighbors as a reference for risk perception and contact adaptation. In future work, we aim to study the impact of behavioral responses driven by disease information collected over varying neighborhood levels.

Finally, our findings reveal a crucial insight into the dynamics of adaptive behavior in epidemic networks. Specifically, our results highlight the public health challenge of identifying suitable individual-level incentives to prompt timely responses at the population level, thereby containing epidemics.
While centralized mandates could foster collective behavioral responses, providing appropriate incentives leads individual decision-making to align with the social optimum~\cite{crabtree2024influential,Traulsen23}.
%
Our results suggest that early and widespread moderate behavioral changes are effective at mitigating the epidemic. However, providing the right incentive to induce appropriate behavioral responses would be challenging.
These findings underscore the potential for targeted behavioral interventions to significantly influence disease dynamics, offering a powerful tool for public health strategies in managing infectious diseases.


\section{Methods}

Our modeling framework assumes that economic productivity depends exclusively on social interaction across edges, over which the disease is transmitted. We model a contagion process on an undirected network, where the network's topology changes over time due to the behavioral adaptations of its nodes at each time step.
Behavioral responses are modeled via a set of Markov Decision Processes (MDPs), where the node-specific optimization process depends on both $(i)$ the number of direct neighbors each susceptible node has in the baseline network structure, and $(ii)$ the locally perceived disease prevalence, the number of infectious neighbors in the $n$-step neighborhood.
We model node-specific decision-making processes by allowing susceptible nodes to randomly remove infectious edges from their neighborhood. The number of infectious edges to drop is determined by an optimization process that privately optimizes the risk of infection and the expected economic benefits derived from social interactions.
The node-specific decision process at each time step uses a projection of the current neighborhood's state during the planning horizon. Optimal decision problems are solved by assuming that the neighbors' health classes at time $t$ remain constant over the planning horizon.

\subsection{Homogeneous adaptive behavior}
Our baseline experiments assume the epidemic starts with a single infectious node, randomly selected from an Erd\"{o}s–R\'{e}nyi graph~\cite{Erdos1960}, composed of $N=500$ nodes, with average nodes' degree $\phi=25$. 

Moreover, we assume all the nodes exhibit a homogeneous sensitivity to infection risk ($\nu=0.05$), and we assign each susceptible node a time-invariant planning horizon ($\tau$). Susceptible nodes are assumed to use a constant projection of the system's state over the planning horizon to compute their expected utilities.

Our framework can generally be extended to account for the risk of infection in the $n$-step neighborhood. Here, we explore where risk is assessed in the $1$-step neighborhood. Furthermore, our framework would incorporate non-linear projections of the future system's state over the planning horizon.

\subsection{Disease transmission model}

The population of interest is determined by a network structure $M$ over which the contagion process occurs. Nodes are assumed to exhibit the health status of Susceptible ($S$), Infected ($I$), or recovered ($R$). We assume the disease is transmitted via the network edges, where $\beta\in [0,1]$ is the transmission probability from an infected $I$ to a susceptible node $S$ along a single edge at each time step.




\subsection{Modeling adaptive human behavior}

We model the behavioral responses of susceptible nodes by dropping edges towards infectious neighbors. At each time step, a susceptible node chooses the number of edges to drop based on the balance between the risk of infection and the benefit of maintaining social interactions.

The edges to drop are randomly chosen among those towards infectious neighbors. We assume that different behavioral responses occur among nodes in distinct health classes, but nodes with similar health statuses follow the same behavioral principles. The node-specific behavioral responses are driven by the health class-specific utility and the node's infection risk at each time step.

The {\em adaptive network structure} produced by the nodes' behavioral responses influences the path of the epidemic and hence future transmission risks. Each node is assumed to be aware of its neighbors' health status (the neighborhood disease prevalence), which is used to infer infection risks and to project the net benefits of social interactions over the planning horizon.

The projection of the future system's state assumes the node distribution among health classes and their respective current social structure (the network structure) remains constant over the planning horizon. The risk of infection for a susceptible node $n_kS$ depends on its edges $e_k$ and the infected neighbors. Removing these edges reduces the node's infection risk but also constitutes a cost. Thus, behavioral adaptation balances feedback between the epidemiological and socio-economic systems.

Each node tends to maximize its expected utility, seeking to maintain social interactions subject to the dynamics of the epidemic. We determine the node-specific optimal behavioral response at each time step by finding the number of edges that maximize its expected utility $V_t(n_k)$ in each of the potential future health states $h\in \{S,I,R\}$, over a given planning horizon, $\tau_k$.

The expected utility $V_t(n_k)$ comprises the potential benefit obtained by making the optimal choices at each future time step during the planning horizon. The node-specific expected utility comprises the immediate net benefits of social interaction (which depend on the node's and its neighbors' health status) and the expected net benefits of future social interactions (which rely on the potential future health states and the corresponding transition probabilities).

Notice that while the node-specific local prevalence depends on the network structure, the potential future health states and the transition probabilities depend on the disease progression model.


Let the node $n_k$ get utility of maintaining $e$ edges at time $t$ ($e_{k,t}$). Utility is described by a concave single-peaked utility function $u_{k,t}=u(e_{k,t})$. Each node obtains a positive marginal net benefit of increasing edges up to $e_k^*$, the node-specific maximum number of edges, determined by the baseline network structure.

Following the work by \cite{morin2013sir}, we assume the node $n_k$ utility function of the particular form $u_{k,t}=\left(b e_{k,t}^{h}-(e_{k,t}^{h})^ 2\right)^{\nu}$, where $b=2 e_k^*$ so that $e_k^*$ produce the maximum immediate utility, $\nu$ is the utility function shape parameter, and $e_{k,t}^{h}$ is the number of edges of a node $k$ with health status $h$ at time $t$.

We assume each node gets a similar per-edge utility regardless of health status, except for infected nodes, which get no utility during their infectious period. We formalize the node-specific optimization problem via a system of Bellman's equations, which are numerically solved at each time step using dynamic programming methods.

\subsection{Susceptible nodes adaptive behavior}

At each time step, each susceptible node chooses the optimal number of edges that maximizes its expected utility over the planning horizon $[t,t+\tau]$.

The optimal number of edges at time $t$ is computed by weighing the current and expected future benefits of social interactions against the risk of infection and potential transitions between health states.

We model the set of optimization problems as dynamic programming problems, whose solutions generate node-specific, privately optimal neighborhood structures \cite{fenichel2011adaptive,morin2013sir,perrings2014merging}.

The dynamic programming problem by which susceptible nodes $n_k^S$ assess their optimal neighborhood structure at time $t$ (the optimal number of edges $e_{k,t}^*$ with which to engage daily) is mathematically formalized by Bellman's equation
\begin{align} \label{eqn:bellman_snodes}
&V_{t}(n_{k,t}^S) = \nonumber \\ &\max_{e_{k,t}^S}\biggl\{u(S,e_{k,t}^S)+\delta[(1-P^I)V_{t+1}(n_{k,t+1}^S)+P^I V_{t+1}(n_{k,t+1}^I) ] \biggr\}
\end{align}

where $V_{t}(n_{k,t}^S)$ is the expected utility of the susceptible node $n_{k}$, at time $t$. Here we define $V_{t+1}(n_{k,t}^S)$ (resp. $V_{t+1}(n_{k,t}^I)$) as the expected utility while being susceptible (resp. infected) at time $t+1$, and
\begin{equation}
    P^I=1-(1-\beta)^{e_{k,t}},
\end{equation}
is the node-specific probability of infection at time $t$, if the node has $e_{k,t}$ edges connecting it to infected neighbors at time $t$.

Equation \eqref{eqn:bellman_snodes} formalizes the maximization problem of susceptible nodes, accounting for the immediate utility $u(S,e_{k,t}^S)$, plus the expected future utility discounted at a rate $\delta$. The expected future utility of susceptible nodes accounts for the expected utility of remaining susceptible at time $t+1$ with probability $1-P^I$, and the expected utility of becoming infected at time $t+1$ with probability $P^I$. The optimization process yields an optimal decision $e_{k,t}^*$ for the number of edges each susceptible individual should select, at time $t$. Each day, the nodes will, therefore, remove edges to infected neighbors to maintain the optimal number of contacts derived from the optimization process. The corresponding formula for $V_{t}(n_{k,t}^I)$ is provided in the appendix for completeness, although infected individuals do not undergo a decision-making process, as we assume infected nodes have no incentive to reduce their social activity level by dropping edges.

\section*{Acknowledgments}
We thank the University of Costa Rica for its support. NSF DMS-2327710, DTRA HDTRA120F0017, NSF CCF-1918656.


\section*{Appendix}

In the Supplementary document, we provide the following.

\begin{itemize}
    \item[(i)] A survey on related literature regarding network models in epidemics and their adaptation techniques. We compare the traditional phenomenological approaches with the purely adaptive method from this article.
    \item[(ii)] Expand on some mathematical details of the model methods introduced in the main article.
    \item[(iii)] Results for adaptive framework model using different structures for the underlying network. We explore other network models, such as the small-world model and the Albert-Barabási model, to examine the robustness of the properties of the adaptive framework discussed in the main article.
    \item[(iv)] Sensitivity of model results to network topological parameters such as network connectivity. We examine the effect of these network parameters on the epidemiological behavior of the disease, as well as the severity of the adaptive process.
    \item[(v)] Comparison of the adaptive network model with the adaptive mean-field formulation. The adaptive framework has so far been formulated for mean-field systems. We explore the differences and advantages that the network implementation brings to this methodology.
    \item[(vi)] An application of the adaptive network model to the study of network layers. The framework proposed in our main article can be further applied to many network scenarios. We inspect the use of multiple contact layers and the effect of different adaptation processes. This extends the capabilities of the adaptive framework to capture even more local information.
\end{itemize}

\section*{Related work}

The use of networks for epidemic modeling has an extensive literature which spans deep mathematical analysis, a variety of algorithms and modeling approaches, and a myriad of applications, \cite{Bansal10,Bansal07,Kiss17,Newman2010,Pastor15}. A detailed overview of the origins and development of network modeling for epidemics can be found in \cite{Keeling2005}. Classical network models compute the individual probabilities of infection and recovery over static network topologies. This formulation soon demonstrated the advantage of using networks to represent the dynamics of infectious diseases more accurately. The primary contribution of the network approach is the influence that network topologies have on disease spread behavior. In \cite{albert2000error}, the relationship between network topology and survivability is demonstrated, whereas \cite{GATES201511} shows that network topology manipulation has an impact on disease transmission dynamics. A study of different network topologies and their impact on epidemic behavior can be found in \cite{leventhal2015evolution}, where the different analytical properties of the network (such as node degrees and density) are shown to affect the epidemic spread significantly, both in theoretical and empirical networks.

Although network models are abundant, we focus on the attempts made to model adaptive behavior in network models. The main classification made for these two types of models, as mentioned in our introduction, consists of two types: evolving and adaptive networks, according to \cite{Guo13}. In the majority of these models, the strategy to perform a node contact adaptation relies on processes of rewiring and/or edge deletion, which follows, in most cases, a phenomenological approach, based mainly on the global status of the epidemics or based on some centralized formula or algorithm proposed at the method level, but not primarily based on individual level attributes or qualities of each node. We now explore some of these formulations for adaptive network modeling. The process of selecting edges to be rewired or broken typically involves probabilistic decisions. In \cite{Zhu2019}, a probabilistic approach is taken to delete connections to infected nodes and replace them with uninfected ones in times of undesired infected probabilities. In \cite{Shivakumar12}, the probability of adding (or cutting) a connection depends on a sensitivity parameter for the corresponding node and also on its degree. However, the decision is based only upon a stochastic process and not a particular optimization process at the node level. In \cite{RISAUGUSMAN200952}, rewiring can be done by both susceptible and infected individuals, where a contact is dropped and a new edge is formed. The node rewiring threshold is also computed depending on node degrees, but the new node is chosen randomly: a node does not know the health status of its new connection; this yields interesting results in terms of disease suppression. On the other hand, more general random link activation-deletion models (RLAD) are proposed in \cite{Kiss2012}, which creates a dynamic network structure, coupled with an SIS dynamics; the purpose of this approach is to offer a benchmark for full simulations. A general explanation of the basic rewiring process can be found in \cite{DONG2015169}, where it is shown that with low link breakage thresholds, the rewiring strategy cannot always contain the epidemic prevalence. In \cite{Ball2022} various possibilities for the rewiring process are studied in full mathematical depth: (i) rewiring to susceptibles only, (ii) rewiring to only recovered individuals or (iii) rewiring to non-infected individuals.

Some models further incorporate the rewiring strategy into more complex scenarios, for example, in \cite{vazquez2016rescue} a two layer network model is proposed with linkage nodes between both layers; these links between network layers are rewired at random to avoid contacts between susceptible and infected nodes at the inter-layer level which results in possible of effective decoupling, allowing for the spread of the epidemic as a whole, but not on each layer separately, resulting in a weak epidemic state. In some other models, the edge creation or deletion can be obtained by means of stochastic processes. For example, in \cite{rocha2013bursts}, a stochastic approach is used to theoretically model bursts of contacts in a network, thus generating heterogeneous contact patterns which generally cause earlier and larger epidemic peaks. 

In most of these network models, the individual economic gain factor is not incorporated into the methodology, thus only focusing on the epidemiological attributes. \cite{Maharaj12} uses a method to perform link rewiring based on a spatial approach, but also employs an individual risk parameter. The economic repercussions of each decision are studied by assigning an economic cost to the loss of contacts. The risk parameter is used to compute the reduction in the size of the immediate awareness neighborhood for each node, using the formula $N_{\text{new}}= N_0(1-\Theta^{\alpha})$, where $N_0$ is the current neighborhood size, $\Theta$ is the current infected density in said neighborhood, $\alpha$ is the risk parameter and $N_{\text{new}}$ is the new size of the neighborhood after edge rewiring or deletion. Although this approach incorporates the sensitivity of individual attributes into the rewiring process, the decision is based on a proportional and spatial consideration, rather than a specific optimization process that accurately captures the agent's risk aversion and projection of the current state. Here is where the adaptive method proposed in this article offers an alternative to capture the risk attributes in a spirit of individual decision and cost comparison.

\section*{Further method details}

\begin{table}[ht!]
\tiny
    \centering
    \begin{tabular}{|c|l|l|}
        \hline
         Parameter & Description & Value \\ \hline
         $\beta$ & Per-edge transmission probability & 0.5 (day$^-1$) \\ \hline
         $\gamma$ & Probability of recovery for infected nodes & 0.4 (day$^-1$) \\ \hline
         $\delta$ & Discount factor & $5\%$ yearly (unitless) \\ \hline
         $\nu$ & Risk sensitivity & $0.05$ (unitless) \\ \hline
         $\tau$ & Planning horizon & 7 (days) \\ \hline
         $e_{k,t}$ & Number of edges of the $k$-th node at time $t$ & Varies (unitless) \\ \hline
         $u(h,e_{k,t}^h)$ & Immediate utility of node $n_k$ with health state $h\in\{S,I,R\}$ & Varies (unitless) \\ \hline
         $V_t(n_{k,t}^h)$ & Expected utility of node $n_k$ with health state $h\in\{S,I,R\}$, at time $t$ & Varies (unitless) \\ \hline
    \end{tabular}
    \caption{Model parameters and variables.}
    \label{tab:modelparams}
\end{table}

\subsection*{Transmission probability computation}

The per-edge transmission probability $(\beta)$ and the number of infected neighbors determine the overall infection probability for a susceptible node at a single time step. In general, the transmission probability of a typical susceptible node is given by

\begin{equation}\label{eqn:prob_infection}
   P^{SI}= 1-(1-\beta)^{n_{inf}}
\end{equation}
where $n_{inf}:=$ number of infected neighbours. On the other hand, infected nodes are assumed to recover with probability $\gamma\in[0,1]$ at each time step.

\subsection*{Infected and recovered nodes behavior}

To solve the optimization problem for susceptible nodes \eqref{eqn:bellman_snodes}, it is required to know the expected utility of infected nodes.

We assume infected nodes have no incentive to reduce their social activity level by dropping edges.
The edge reduction is produced by susceptible nodes dropping the edges towards infectious ones.
The expected utility computed by a susceptible node while potentially becoming infectious in the future accounts for all connected edges at time $t$. In this regard, the expected utility does not incorporate the potential behavioral decisions of other susceptible nodes, which would drop the edge towards the node while infectious.

Finally, we let recovered nodes maintain the neighborhood structure that maximizes the net benefits of social interactions. The expected utility of infected nodes is given by Bellman's equation \eqref{eqn:bellman_inodes}

\begin{align} \label{eqn:bellman_inodes}
& V_t(n_{k,t}^I) = u(I,e_{k,t}^I)+ \delta\biggl[(1-P^R)V_{t+1}(n_{k,t+1}^I)\,+\,P^R V_{t+1}(n_{k,t+1}^R)\biggr],
\end{align}
where $P^R=\gamma$ is the probability of recovery at each time step.

The expected utility of infected individuals accounts for the immediate utility of being infected and the discounted future expected utility. The infected expected future utility accounts for the expected utility of remaining infected at time $t+1$ with probability $1-P^R$ and the expected utility of recovery at time $t+1$ with probability $P^R$.


Finally, to solve \eqref{eqn:bellman_inodes}, it is required to know the expected utility of recovered nodes. Since our disease progression model assumes permanent immunity of recovered individuals, we assume there is no incentive for recovered individuals to seek strategic behavior. Thus, we let recovered nodes maintain the daily neighborhood structure that maximizes the net benefits of social interactions.
given by the equation \eqref{eqn:bellman_rnodes}

\begin{equation} \label{eqn:bellman_rnodes}
V_t(n_{k,t}^R)=u(R,e_{k,t}^R)+\delta[ V_{t+1}(n_{k,t+1}^R)].
\end{equation}

\section*{Sensitivity to network structure and connectivity}

We explore the adaptive algorithm in networks using different network structures. We study three main types of underlying network models: the Erd\"{o}s–R\'{e}nyi graph~\cite{Erdos1960}, the scale-free model \cite{Barabasi99}, and the small world model \cite{Strogatz98}, varying the network definition parameters for each model. Our main goal is to inspect the main behavioral phenomenon discussed in the article with respect to the network topology. We remind the reader that this phenomenon consists of the presence of a day lag between the maximum contact change made at the global level and the corresponding maximum effort made at the individual level (at the susceptible population). Our results indicated that both contact reductions occurred asynchronously, with the maximum edge reduction at the susceptible level occurring later. We present the result of different simulations in which the network topology is modified and the corresponding day lag is computed for each variation. A summary of each network model and its application to the epidemic simulation process can be found in \cite{Keeling2005}.

\begin{figure}[H]
    \centering
    \includegraphics[scale=0.8]{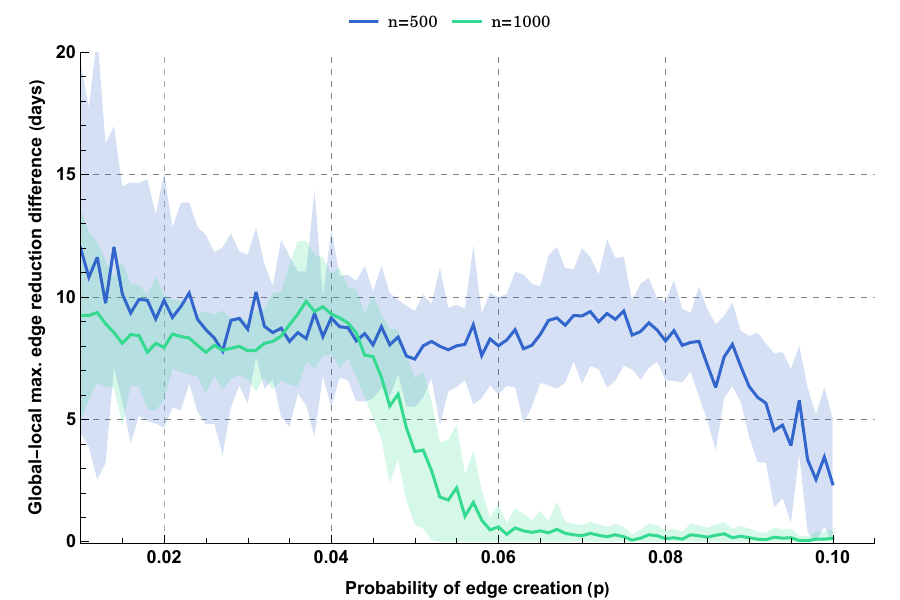}
    \caption{Average maximum behavior reduction day lag difference for a random network simulation, depending on the network size and the probability of edge creation, $p$. Larger values of $p$ will attenuate the adaptive effect, especially for larger populations.}
    \label{fig:connectivityPlot}
\end{figure}

Firstly, we continue to use the Erd\"{o}s–R\'{e}nyi graph, as in the main article. This graph model carries out the edge creation process using a defined probability of edge creation, $p$. Figure \ref{fig:connectivityPlot} shows the average day lag between the maximum contact reductions at the local and global levels for each $p$ in an adaptive model with an underlying Erd\"{o}s–R\'{e}nyi network of $n$ nodes. This plot was done keeping the adaptive framework parameters constant, using a planning horizon of $T=7$ days and a utility shape parameter of $\nu = 0.05$. We see that it is only up to a point (approximately $p=0.08$) that the increase in the edge connections reduces the day lag. This indicates that the local responses become \textit{synchronized} with the global behavior when the network becomes more connected. This behavior will be observed for other networks as well, as it will be seen shortly. 

\begin{figure}[ht!]
  \centering
  \begin{subfigure}[b]{0.48\linewidth}
    \centering
    \includegraphics[width=\linewidth]{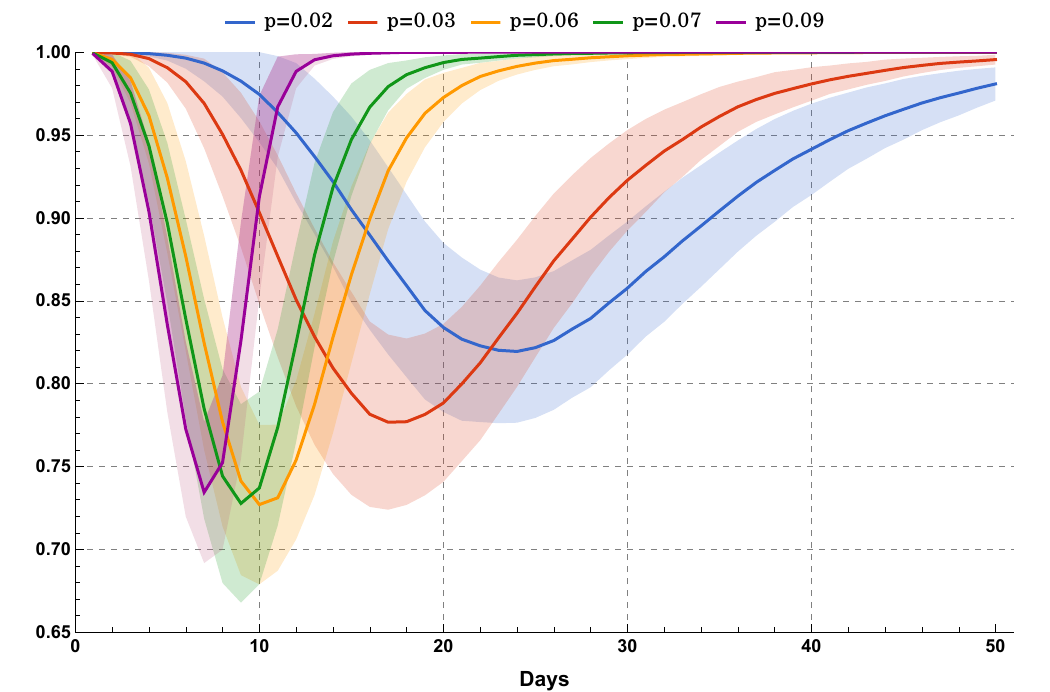}
    \caption{Average network edge reduction.}
    \label{fig:AvgEdgeRedErdos_A}
  \end{subfigure}
  \hfill
  \begin{subfigure}[b]{0.48\linewidth}
    \centering
    \includegraphics[width=\linewidth]{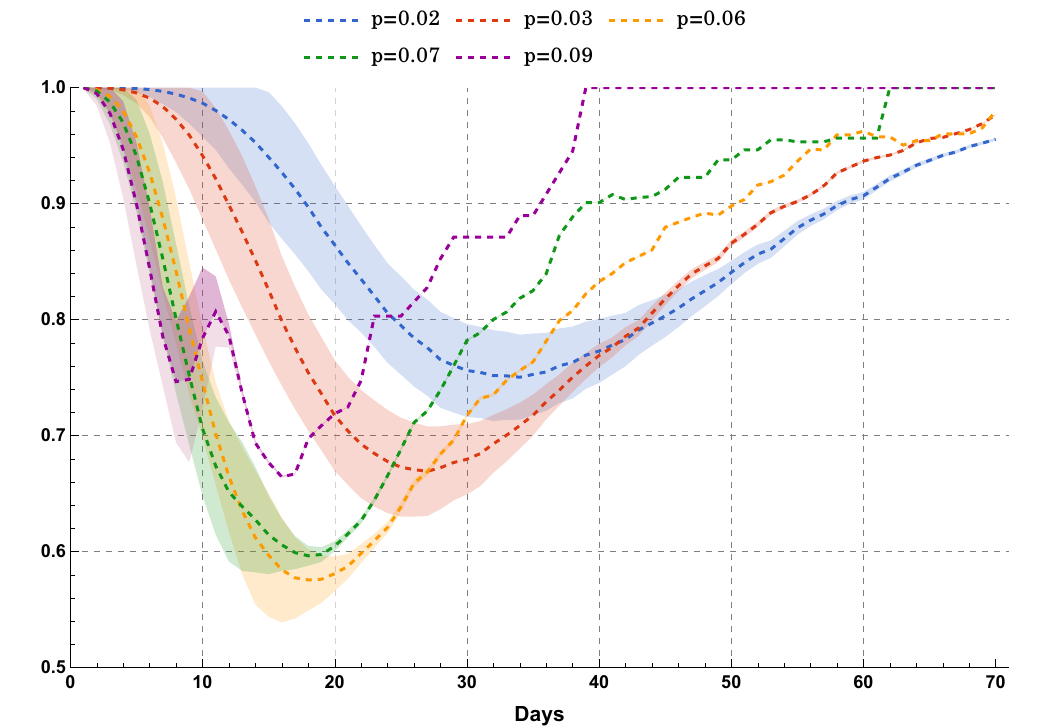}
    \caption{Average individual (susceptible) edge reduction.}
    \label{fig:AvgEdgeRedErdos_B}
  \end{subfigure}
  \caption{(a) Average network edge reduction on Erd\H{o}s–R\'{e}nyi graphs ($n=500$) as function of edge probability $p$. (b) Average susceptible edge reduction under the same simulation.}
  \label{fig:AvgEdgeRedErdosPlot}
\end{figure}

In Figure \ref{fig:AvgEdgeRedErdosPlot}, we offer a further examination of the connectivity parameter, $p$, and the effect it has on the network topology modification. Increasing the connectivity leads to greater stress on contact change at the local and global levels. However, higher values of the edge creation probability $p$ will reverse this phenomenon. We observe the irregular behavior of the susceptible contact modification when $p$ is large.


\begin{figure}[H]
    \centering
    \includegraphics[scale=0.75]{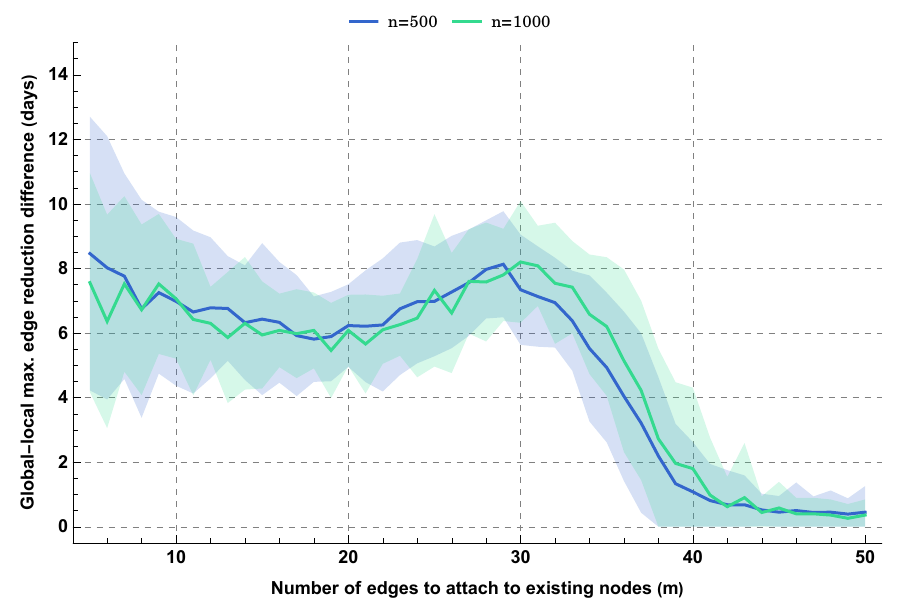}
    \caption{Average maximum behavior reduction day lag difference, using an Albert-Barabasi model for the underlying network, for each value of the parameter $m$ of this network model and varying the population size. Unlike the case with the random network model, population size has little impact on the day lag difference.}
    \label{fig:BarabasiAlbertPlot}
\end{figure}

When using the Albert-Barabasi model, \cite{Barabasi99}, new nodes are progressively added to the graph by means of attaching an average of $m$ nodes to existing nodes with preferably high degree. Experimenting with this network model as the underlying graph for the adaptive framework, we find that the main day lag phenomenon of the article is still present. Like in the previous model, the edge connectivity plays a role in the magnitude of this day lag. In figure \ref{fig:BarabasiAlbertPlot} see that when increasing the parameter $m$, the local and global behavior become similar as well. For this network model we observe that the population size ($n$) has a smaller overall effect.

\begin{figure}[ht!]
  \centering
  \begin{subfigure}[b]{0.48\linewidth}
    \centering
    \includegraphics[width=\linewidth]{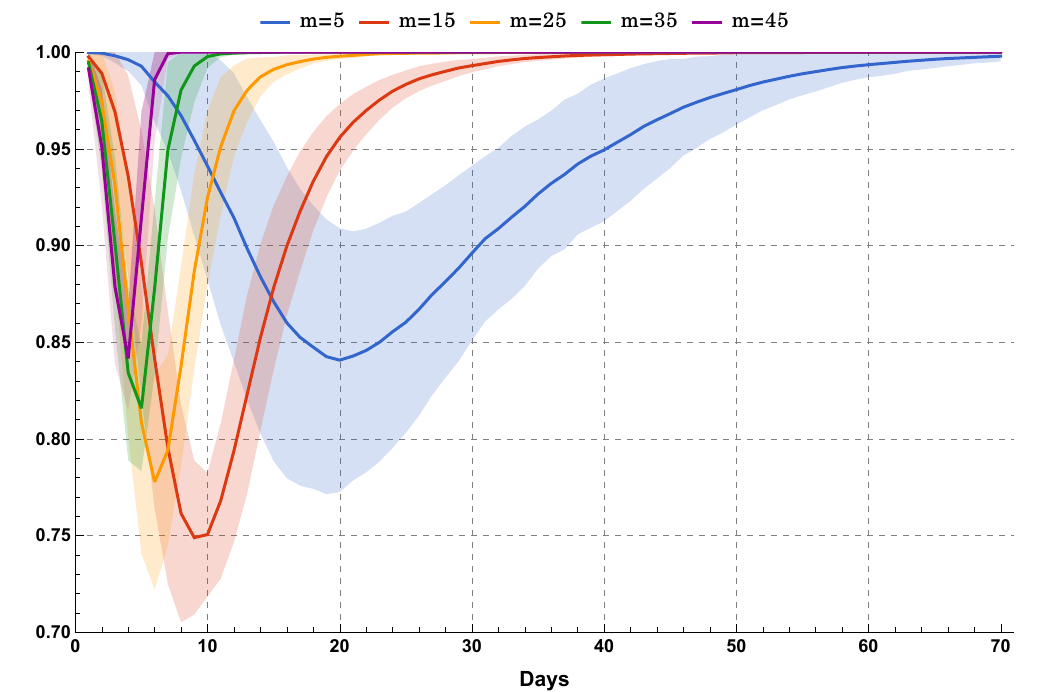}
    \caption{Average network edge reduction.}
    \label{fig:AvgAlbertBarabasi_A}
  \end{subfigure}
  \hfill
  \begin{subfigure}[b]{0.48\linewidth}
    \centering
    \includegraphics[width=\linewidth]{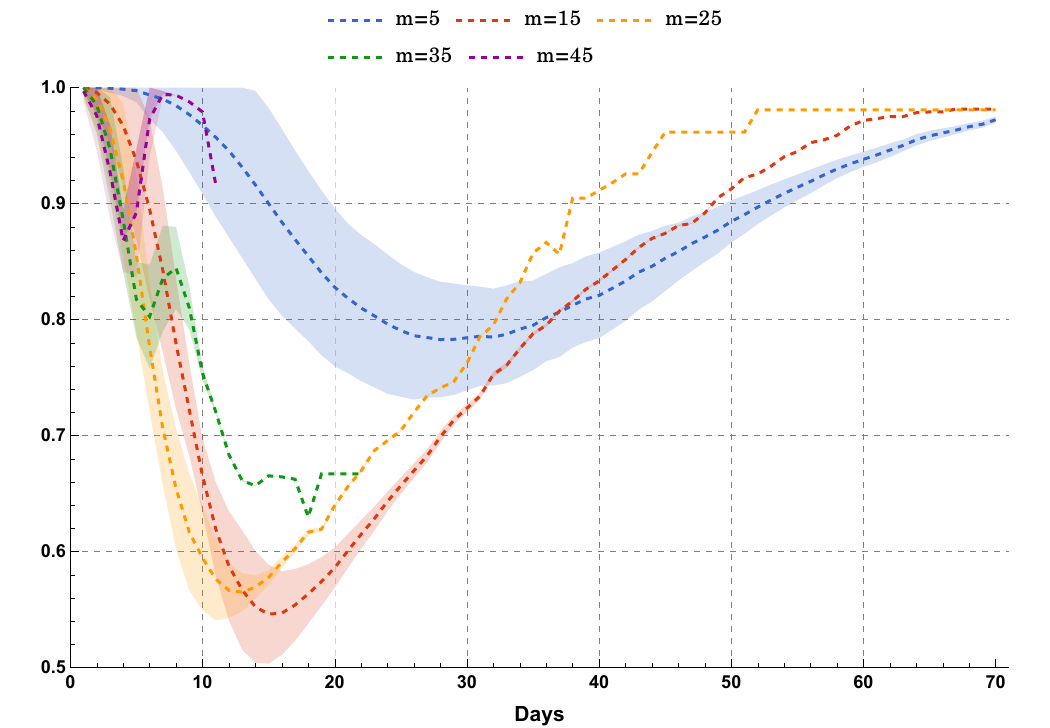}
    \caption{Average individual (susceptible) edge reduction.}
    \label{fig:AvgAlbertBarabasi_B}
  \end{subfigure}
  \caption{(a) Average network edge reduction using an underlying Albert–Barabási network with $n=500$ nodes, depending on the edge creation probability $p$. (b) Average susceptible reduction under the same simulation.}
  \label{fig:AvgAlbertBarabasiPlot}
\end{figure}

Figure \ref{fig:AvgAlbertBarabasiPlot} shows a similar behavior of the average network and individual edge reduction, where an initial increase of the parameter $m$ leads to more drastic edge reductions, up to a point where the connectivity of the network ensures a rapid much more rapid spread of the disease and therefore diminishes the overall edge reduction that results from the adaptive decision process. Note that edge reduction curves for susceptibles stop when there are no more susceptible individuals in the population.


\begin{figure}[H]
    \centering
    \includegraphics[scale=0.65]{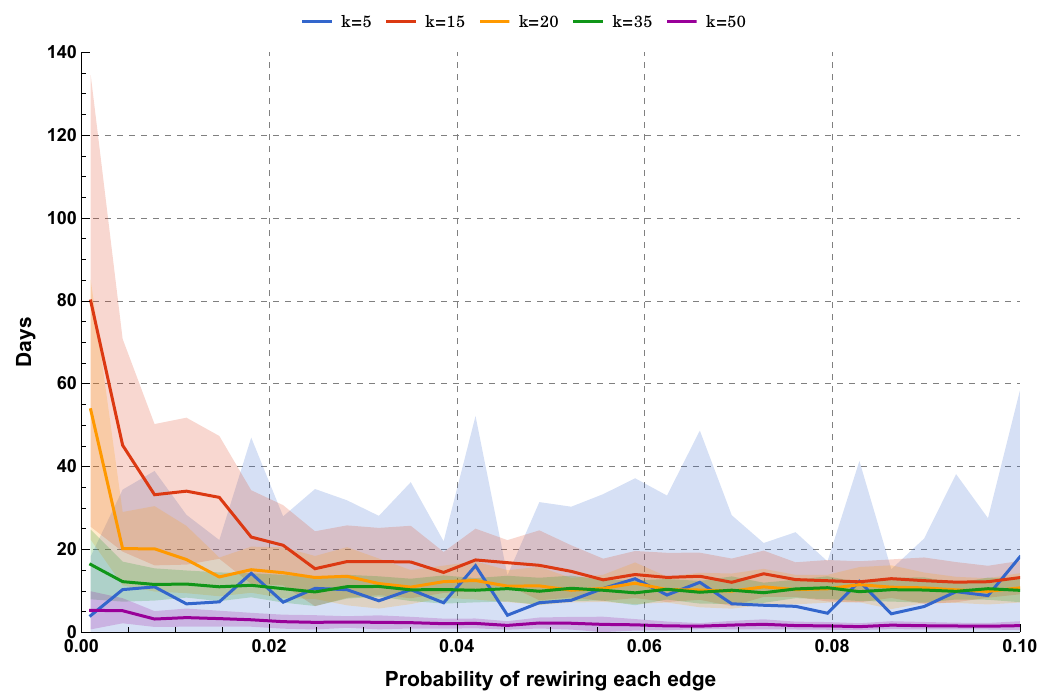}
    \caption{Average maximum behavior reduction day lag difference, using a small world model for the underlying network with parameters $k$, the number of neighbors in the base ring topology, and $p$, the probability of rewiring each node.}
    \label{fig:SmallWorldPlot}
\end{figure}

Finally, we explore the small world model from \cite{Strogatz98}. The basic procedure to create a small world model starts with the creation of ring lattices of $n$ nodes and $k$ edges and then proceeds to perform random edge rewiring with a probability of $p$.  Figure \ref{fig:SmallWorldPlot} displays the average day lag between the maximum edge reductions at the local vs global level, for different values of $k$ and $p$. Increasing both parameters leads to a higher network connectivity, which yields a smaller day lag difference and an approximation of the adaptive framework to the classical static network model.

\begin{figure}[ht!]
  \centering
  \begin{subfigure}[b]{0.48\linewidth}
    \centering
    \includegraphics[width=\linewidth]{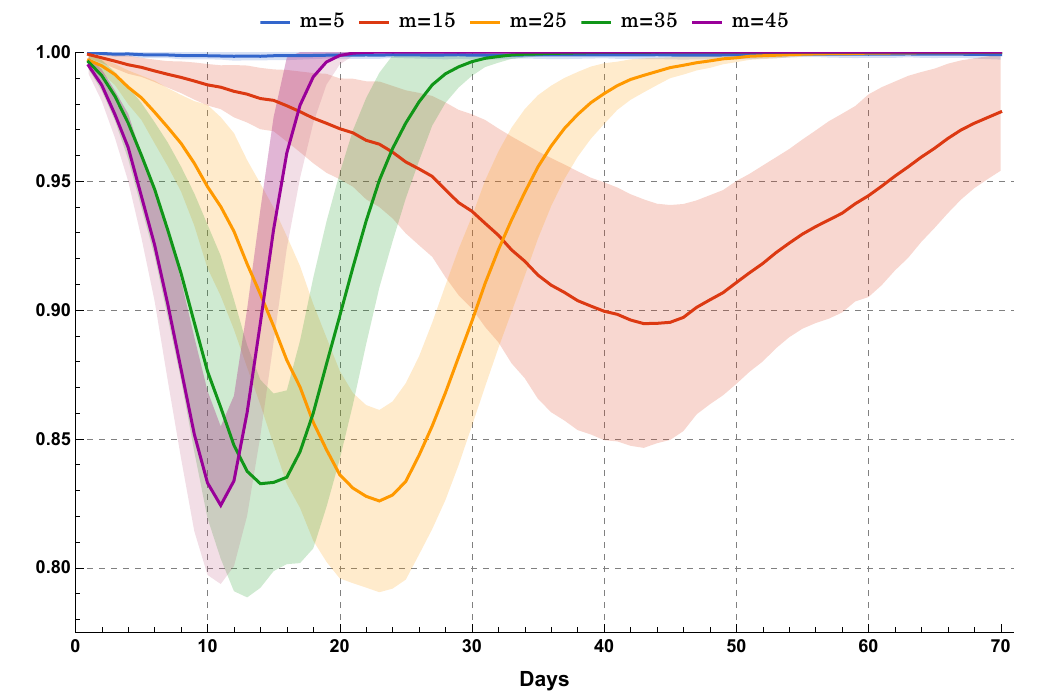}
    \caption{Average network edge reduction.}
    \label{fig:AvgSW_A}
  \end{subfigure}
  \hfill
  \begin{subfigure}[b]{0.48\linewidth}
    \centering
    \includegraphics[width=\linewidth]{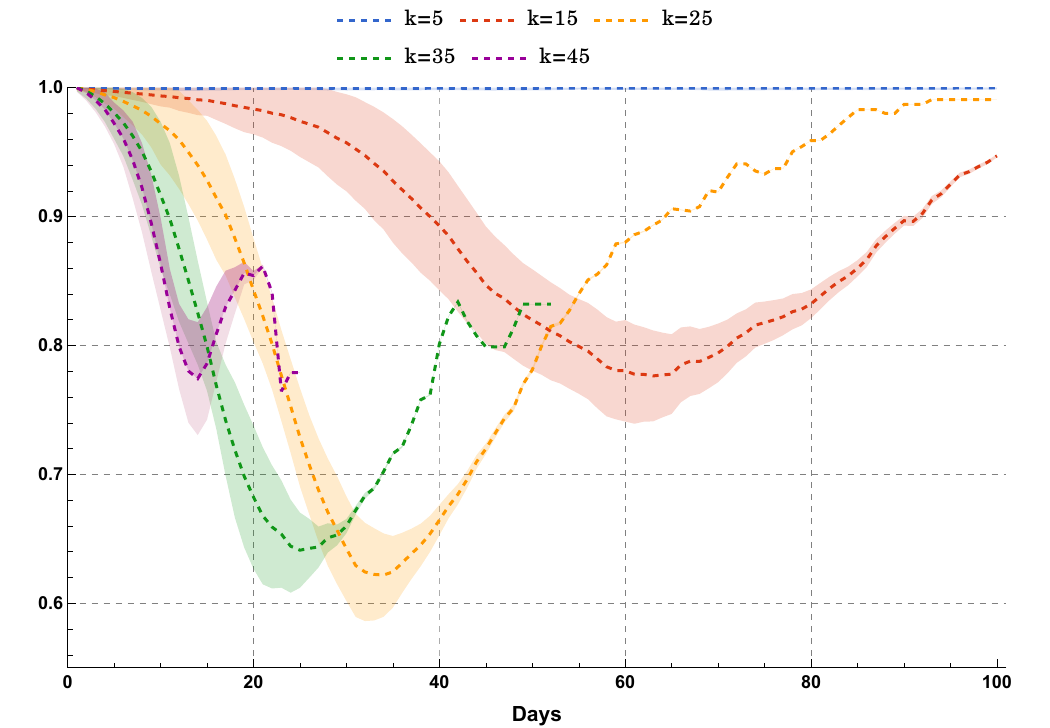}
    \caption{Average individual (susceptible) edge reduction.}
    \label{fig:AvgSW_B}
  \end{subfigure}
  \caption{(a) Average network edge reduction using a Small-World network with $n=500$ nodes and rewiring probability $p=0.023$, varying with the number of neighbors $k$ in the initial ring. (b) Average susceptible edge reduction in the same simulation.}
  \label{fig:AvgSWPlot}
\end{figure}

\begin{figure}[ht!]
  \centering
  \begin{subfigure}[b]{0.48\linewidth}
    \centering
    \includegraphics[width=\linewidth]{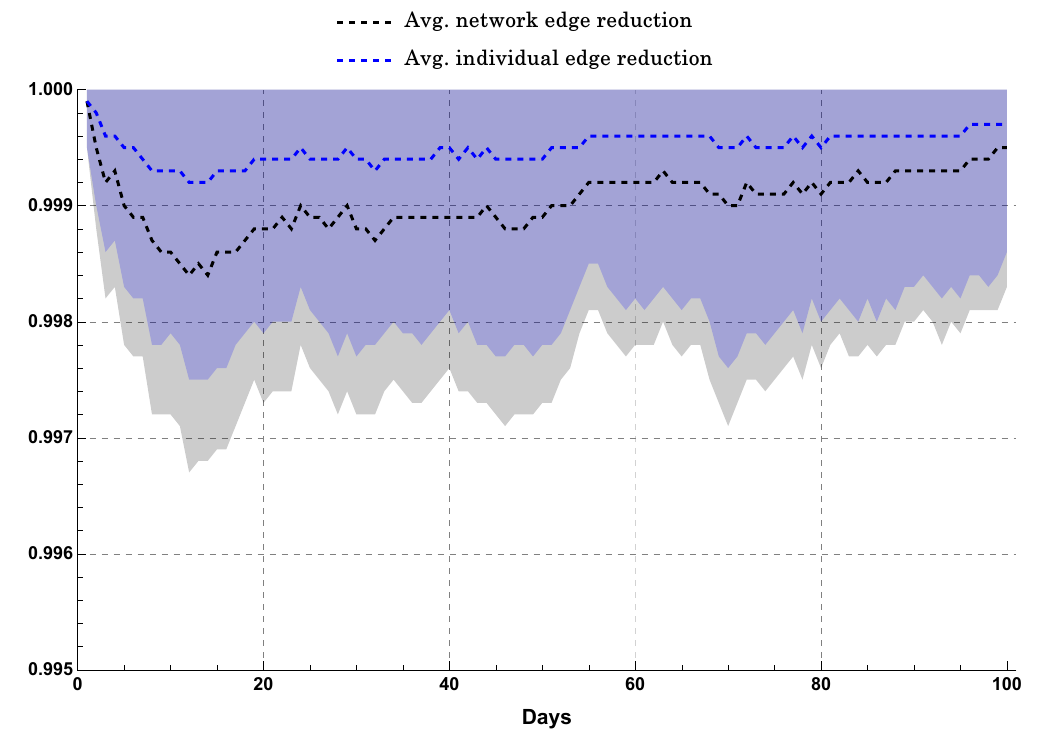}
    \caption{Average network and individual edge reduction.}
    \label{fig:SmallWorldK5_A}
  \end{subfigure}
  \hfill
  \begin{subfigure}[b]{0.48\linewidth}
    \centering
    \includegraphics[width=\linewidth]{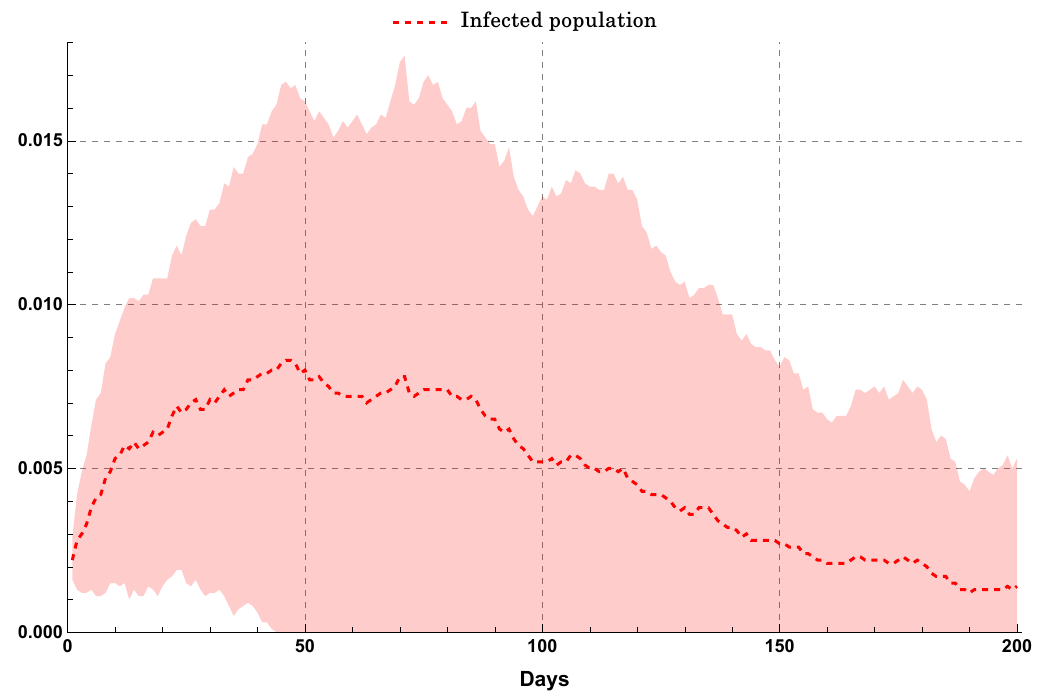}
    \caption{Average infected population.}
    \label{fig:SmallWorldK5_B}
  \end{subfigure}
  \caption{(a) Average network and individual edge reductions for a Small-World network with a ring topology of $k=5$ neighbors. (b) Corresponding infected proportion over time. This represents an extreme case in which sparse connectivity hinders meaningful behavioral adaptation.}
  \label{fig:SmallWorldPlotK5}
\end{figure}

Figure \ref{fig:AvgSWPlot} displays similar contact reduction plots for a model with an underlying small-world network, using $n=500$ nodes and a probability of rewiring of $p=0.023$. In this model, we note an unusual behavior for specific parameters. We can see a greater variation in plot \ref{fig:AvgSWPlot} for smaller values of $p$ and $k$, a being $k=5$ a notable case. These examples constitute cases in which the network connectivity can be so low that the central phenomenon of asynchronous global/local responses does not take place. In Figure \ref{fig:SmallWorldPlotK5} we display the contact reduction at the global and local levels, as well as the average infected proportion for an underlying small world network with parameters $k=5$ and $p=0.078$.


The underlying network topology is an essential factor in the disease spreading within epidemic network models, \cite{Ganesh05,Meyers07,Perez22}, and we see the corresponding effect in our main results concerning the adaptive framework and its application to a network structure. From Figures \ref{fig:AvgEdgeRedErdosPlot} and \ref{fig:AvgAlbertBarabasiPlot}, we can see that an initial increase in network density induces a greater global adaptive response. However, at some point, the network topology becomes too dense, and this diminishes the effect of the adaptive process, making the network model approach towards a classic network with no contact reduction.

\begin{figure}[ht!]
  \centering
  \begin{subfigure}[b]{0.48\linewidth}
    \centering
    \includegraphics[width=\linewidth]{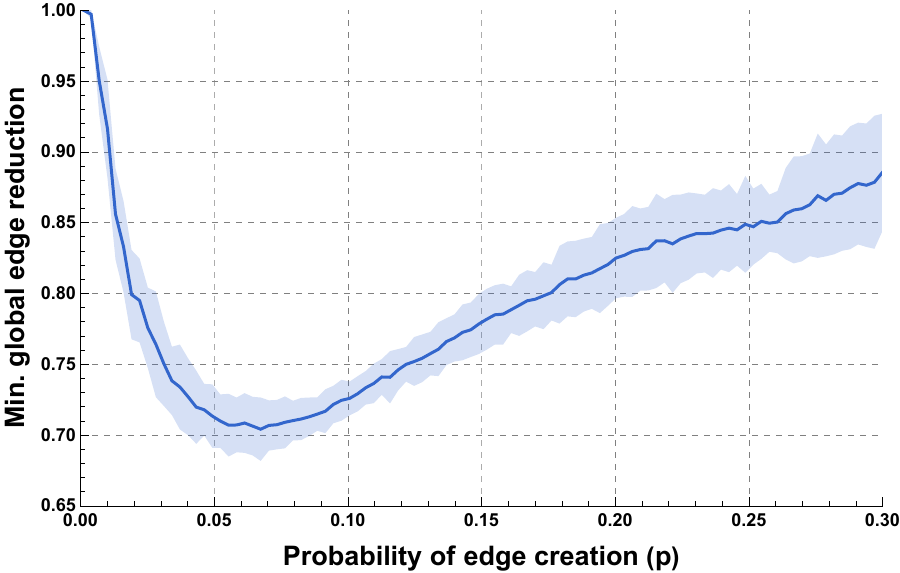}
    \caption{Erd\H{o}s–R\'{e}nyi network.}
    \label{fig:MinPlots_Erdos}
  \end{subfigure}
  \hfill
  \begin{subfigure}[b]{0.48\linewidth}
    \centering
    \includegraphics[width=\linewidth]{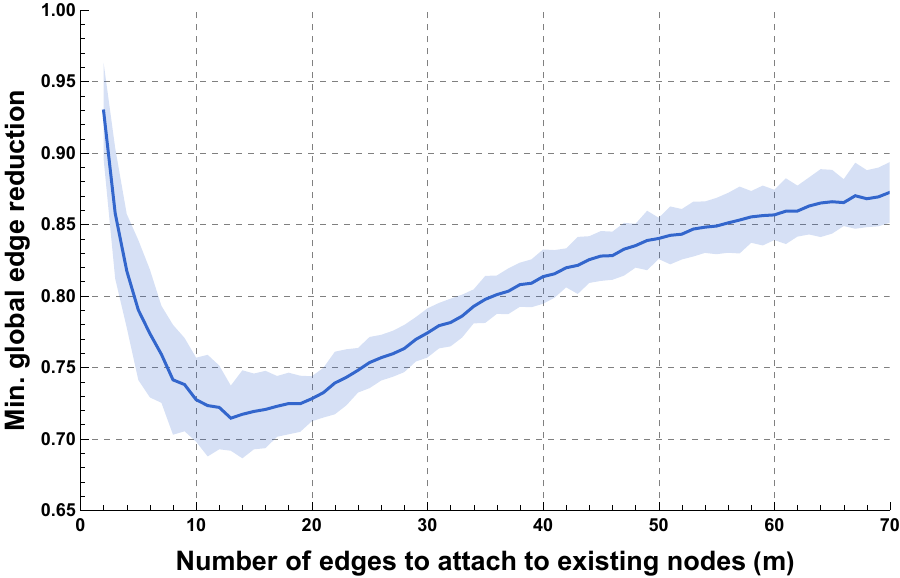}
    \caption{Albert–Barabási network.}
    \label{fig:MinPlots_Albert}
  \end{subfigure}
  \caption{Minimum global edge reduction using (a) an Erd\H{o}s–R\'{e}nyi network model with $n=500$ nodes and varying edge probability $p$, and (b) an Albert–Barabási network model with $n=500$ nodes and varying the average number of edges $m$ each new node forms with existing nodes.}
  \label{fig:MinPlots}
\end{figure}

This phenomenon can be seen in Figure \ref{fig:MinPlots}, where the minimum global edge reduction depends on the topology of the underlying network. For low values of $p$ (in the Erd\"{o}s–R\'{e}nyi model) or $m$ (in the Albert-Barabasi model), this minimum is close to $1$, meaning that the adaptive response is practically null as a result of highly sparse network structures. Augmenting the network's connectivity increases the likelihood of the adaptive decision process, leading to more relevant edge reduction responses. Eventually, the high density becomes a factor that accelerates disease spread, and this means a rapid reduction of susceptibles, neutralizing the adaptive process and resulting in a reduction of the adaptive response.

We remark that these results have been done employing a fixed disease transmission probability, $\beta$. In mean field models, the $\beta$ parameter summarizes disease spreading by accounting for probabilities both for infection and for contacts with infected individuals. In network models, particularly with the use of the adaptive framework, these two processes are separated; the $\beta$ accounts for the probability of being infected by a contact with infected agents, and the likelihood of contacting these individuals is controlled by the network topology. This represents a great advantage of the use of network models, where the network structure and the disease-level parameters play an intertwined role in the epidemic progress. Varying the value of $\beta$ in these simulations will provide different windows for a meaningful adaptive process to take place.


\textbf{Remark.} We point out that all of these results have been computed using a neighborhood of one step in computing the probability of infection. A case could be argued that the adaptive decision could also be based on further levels of neighbors, which could give more options for a fuller adaptive consideration to take place daily. However, using further neighborhood levels to calculate this probability of infection represents a computational problem for which no efficient solution is known (that is, an NP-hard problem), as proved in \cite{SHAPIRO201277}.

\section*{Adaptive behavior on real-world networks}

\begin{figure}[h!]
    \centering
    \includegraphics[width=0.8\linewidth]{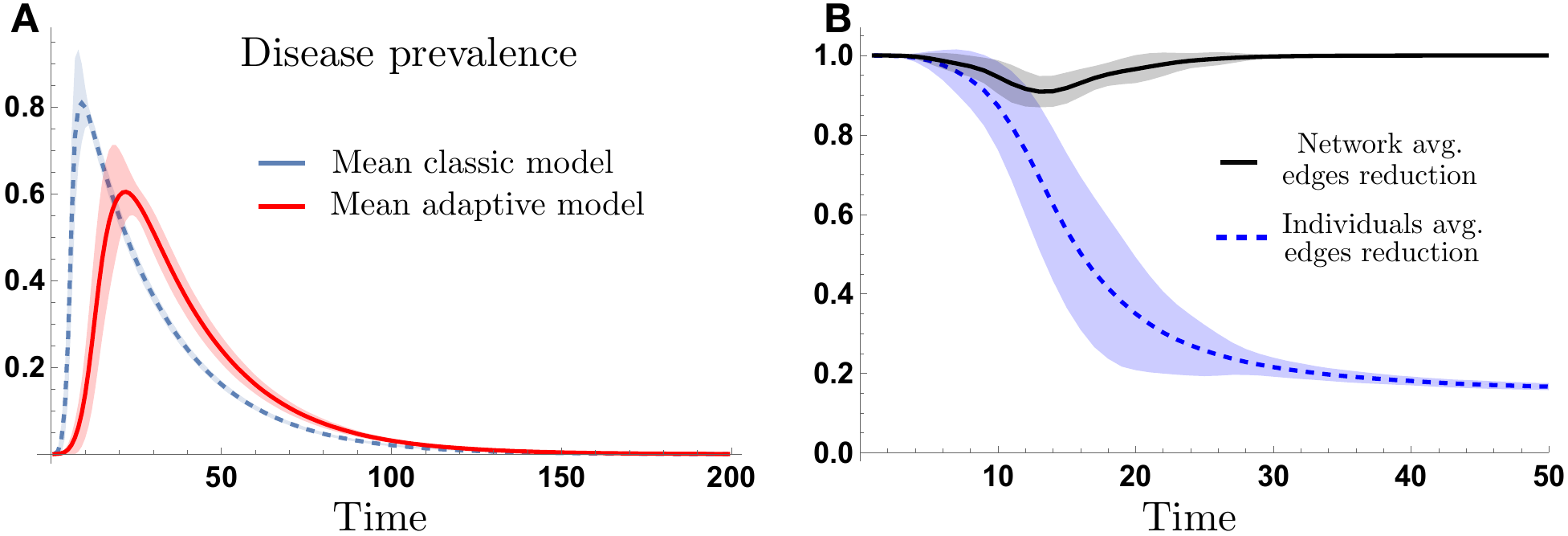}
    \caption{
    {\bf Disease and adaptive behavior dynamics in Manassas network.}
    (Panel) The epidemic dynamics for the classic and the adaptive behavior model in the selected subnetwork of Manassas.
    (Panel B) The normalized average edge reduction at the population and individual scales as the epidemic progresses.
    }
    \label{fig:man_adapt}
\end{figure}

We study the impact of heterogeneous adaptive behaviors on a realistic
social contact network. We use a digital twin of Manassas city, Virginia, social contact network generated by $(i)$ constructing synthetic individuals and locations; $(ii)$ assigning daily activity sequences to each individual; $(iii)$ mapping each activity of each individual to a location with the associated visiting time interval; $(iv)$ deriving edges between people from their physical proximity when they visit the same location at the same time.
The detailed methodology of the construction of this type of digital twin for any county or state level region in the U.S.A. is described in~\cite{barrett2009generation,chen2021prioritizing,eubank2004modelling,Mortveit:20}.
We consider an unweighted subnetwork of the Manassas network of $5,000$ nodes with an average degree of $22$, with the same parameter values used in our main manuscript.
Our simulations in Figure~\ref{fig:man_adapt} show that adaptive behavior significantly modulates the epidemic progression by decreasing and delaying the epidemic peak. Moreover, our results show that the asynchronization between the population and individual scale behavioral responses persists in the subnetwork of the Manassas social contact network. Note that the observed behavioral dynamics are less pronounced compared with our results using synthetic networks. Nevertheless, the magnitude of the asynchronization is modulated by the population's behavioral and epidemic characteristics.

\section*{Comparison with mean field adaptive model}

\begin{figure}[ht!]
  \centering
  \begin{subfigure}[b]{0.48\linewidth}
    \centering
    \includegraphics[width=\linewidth]{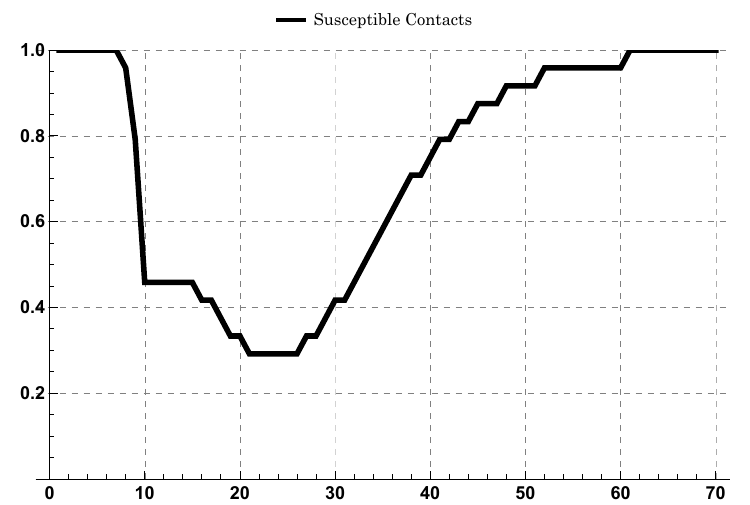}
    \caption{Contact reduction for susceptibles in adaptive mean-field model.}
    \label{fig:adaptive_sir_mf}
  \end{subfigure}
  \hfill
  \begin{subfigure}[b]{0.48\linewidth}
    \centering
    \includegraphics[width=\linewidth]{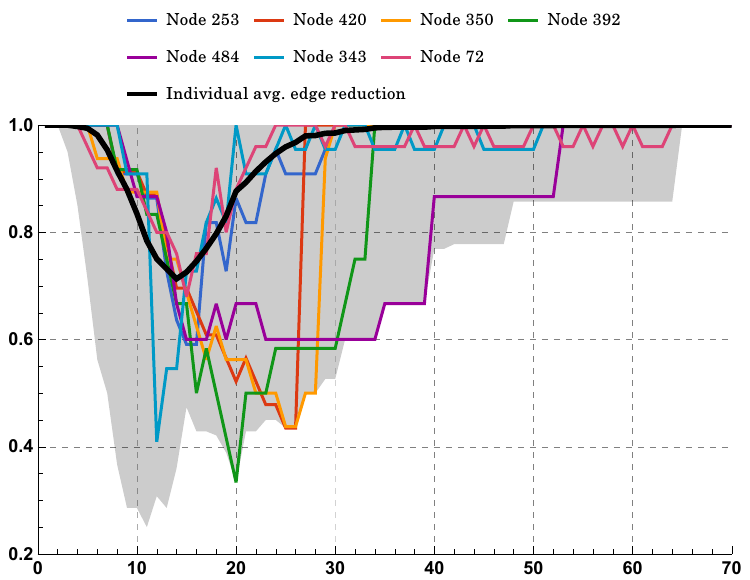}
    \caption{Individual contact reduction in adaptive network model.}
    \label{fig:adaptive_sir_network}
  \end{subfigure}
  \caption{(a) Average contact reduction for susceptibles in an adaptive mean-field model. (b) Sample trajectories of individual contact reduction for seven nodes in a single adaptive network simulation. The black line represents the average edge reduction across all nodes.}
  \label{fig:adaptive_sir_network_comp}
\end{figure}

The adaptive framework has been presented in \cite{fenichel2013economic,fenichel2011adaptive,morin2013sir,perrings2014merging} and has so far been formulated for mean-field models. The contact adaptation occurs for susceptible individuals as an aggregate (as it is the nature of mean field models), resulting in modification of the average contact rate for all susceptible individuals in the population. However, in network models, the information transitions towards the local level, resulting in a specific contact rate adaptation for each individual. The contact modification will depend on features of the neighborhood of each node, such as its size and the number of infected individuals in it. We display a comparison of the adaptive framework for mean-field models and for networks in Figure \ref{fig:adaptive_sir_network_comp}. The reader can observe that whereas the mean-field model only provides contact information for susceptible individuals as an aggregate, the network model provides a specific contact behavior of each individual in the population, highlighting individuals with more extreme contact changes than others. We remark that although the adaptive procedure is carried out only for susceptible individuals in the network at each time step, the resulting edge reduction may also affect non-susceptible nodes.


Neither of these simulations shares the same disease parameters, such as the probabilities of infection and recovery. Clearly, to make a deeper epidemiological comparison between the two models, we would need to inquire about the health metrics that govern disease transmission in both cases, such as the basic reproductive number $R_0$. The purpose of this discussion is mainly to show the structural differences of the contact adaptation process between the mean-field approach and the network model. Besides the difference in complexity and the possibility of modeling more diverse behavior patterns, we note an abrupt global contact decrease in susceptibles as the disease prevalence approaches its maximum. In contrast, the global response in the network model is more measured. This phenomenon is frequent with the adaptive framework implementations at mean field models \cite{Balt2, Balt1, espinoza2021adaptive}. Although panel (b) in Figure \ref{fig:adaptive_sir_network_comp} shows us the existence of nodes with some extreme behavior, we note that the average global behavior is smoother, which is a natural feature of network models.

\section*{Layer information and adaptive networks}

We further explore the advantages of our adaptive framework to incorporate individual decision processes. As we saw in the main article, the adaptive methodology allows us to define risk perceptions at the individual level, which allows us to model different sub-populations within the epidemic development. We found that the global response averages the behavior of each sub-population. The incorporation of node-level information in the adaptive framework provides the possibility to study many locally and economically defined contact phenomena, which are essential parameters in the global disease evolution. In this example, we explore the response towards different contact layers. Not only, the use of multilayered networks offers rich network topology properties, \cite{Bianconi18}, but it also can be used to model complex contact processes, some examples include: the study of reasons for mask usage, \cite{Qiu22}, the modelling of different pathogen strains and their spreading processes, \cite{Sood23}.

In \cite{garcia2022projecting,Sanchez23,SANCHEZ2022100577}, a multilayered network process aims to capture different types of contacts individuals might have, such as family and friends, coworkers, and sporadic contacts, by creating network layers for each contact type. This approach offers a valuable and plausible interpretation of everyday contact dynamics. In network models as these, contact decisions are modelled by having individuals decide their engagements based on uniform distributions predetermined for each layer type. In this case, the adaptive framework can offer a pathway to incorporate an optimal decision process based on the risk perceptions agents have towards contacts of each layer. In the following two examples, we examine simple scenarios to model different contact types and the corresponding risk-informed decision processes that take place at the individual level towards each layer. 

\subsection{Layers as interposed networks} The individual's response to each contact layer can vary. For example, changing contacts with family members or with daily or close acquaintances can be more difficult than with other less important contacts, such as acquaintances or sporadic contacts. This contact group division constitutes a relevant scenario in many epidemiological settings. An early application for mean-field models in the context of sexually transmitted diseases can be found in \cite{HADELER199541}. In the case of network models, there are numerous implementations and theoretical analyses of multilayered networks and their applications in these types of epidemic settings, \cite{Aleta20,Hammoud20,Zhang23}.

\begin{figure}[H]
    \centering
    \begin{tikzpicture}[scale=1, 
    suscnode/.style={person, scale=0.9},
    basenode/.style={person, shirt=black, scale=1},
    infnode/.style={person, shirt=red, skin=green, scale=0.9},]
    

    \filldraw[color=blue!90, top color=blue!20, bottom color=blue!50, very thick] (1,3) -- (2,1) -- (5.982,1.02) -- (4.8,3) --(1,3);

    \filldraw[color=blue, fill=blue!50, thick] (1,3) -- (2,1) -- (6,1) -- (6,0.9) -- (2,0.9) --(1,2.9) --(1,3);

    \filldraw[color=blue, fill=blue!50, thick] (2,1) -- (2,0.9);
    
    \node[basenode] (A1) at (3,2) {};
    \node[suscnode] (B1) at (2.4,2.5) {};
    \node[suscnode] (C1) at (2.3,1.6) {};
    \node[suscnode] (D1) at (3,1.3) {};
    \node[suscnode] (E1) at (3.9,1.7) {};
    \node[infnode] (F1) at (3.6,2.7) {};

    \draw[-, color=black!70] (A1) -- (B1);
    \draw[-, color=black!70] (A1) -- (C1);
    \draw[-, color=black!70] (A1) -- (D1);
    \draw[-, color=black!70] (A1) -- (E1);
    \draw[-, color=black!70] (A1) -- (F1);


    \filldraw[color=orange!90, top color=orange!20, bottom color=orange!50, very thick] (-1,6) -- (0,4) -- (5.982,4.02) -- (4.8,6) -- (-1,6);

    \filldraw[color=orange, fill=orange!50, thick] (-1,6) -- (0,4) -- (6,4) -- (6,3.9) -- (0,3.9) --(-1,5.9) --(-1,6);

    \filldraw[color=orange, fill=orange!50, thick] (0,4) -- (0,3.9);

    \node[basenode] (A2) at (3,5) {};
    \node[infnode] (B2) at (1,5.6) {};
    \node[suscnode] (C2) at (0.5,4.4) {};
    \node[suscnode] (D2) at (5,4.6) {};
    \node[suscnode] (E2) at (4.4,5.7) {};

    \draw[-, color=black!70] (A2) -- (B2);
    \draw[-, color=black!70] (A2) -- (C2);
    \draw[-, color=black!70] (A2) -- (D2);
    \draw[-, color=black!70] (A2) -- (E2);

    \draw[-, color=black!50, dashed] (A1) -- (A2);
    
    \node[text width=3cm] at (7.5,1.8) 
    {Neighbor contacts};

    \node[text width=4cm] at (8,5) 
    {Outer sporadic contacts};

\end{tikzpicture}
\caption{Two types of contact layers for each node: a closely connected network of neighbors and regular contacts, and another broad network of sporadic or irregular contacts. The adaptive framework is performed by the computation of two simultaneous decision processes.}
\label{fig:two_layers_contacts_model}
\end{figure}
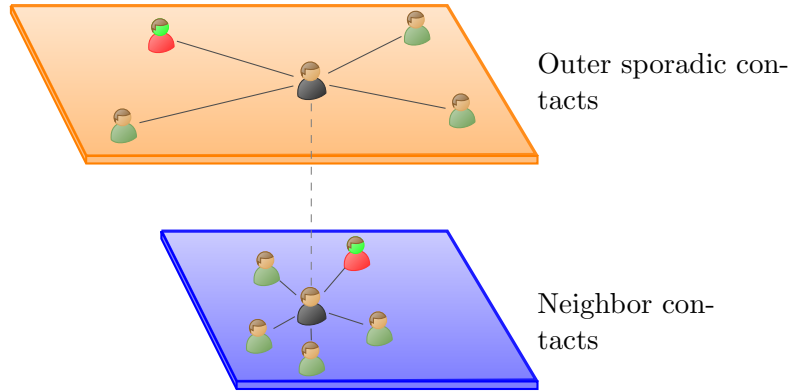

The adaptive framework enables the making of optimal decisions across different layers, each with its specific risk perception. Consider the example of two underlying layers. We model each as a different underlying network, with varying parameters of connectivity, the outer network being a sparser one (see Figure \ref{fig:two_layers_contacts_model} for an illustration of this situation). We perform the adaptive decision process for each network separately, using two utility functions $u_1(\nu_1, \tau_1)$ and $u_2(\nu_2, \tau_2)$, for each network. Given that overall contacts are a combination of both layers, the probability of infection, $P^{SI}= 1-(1-\beta)^{n_{inf}}$, is to be computed using $n_{inf} = n_{inf1} + n_{inf2}$, the number of infected neighbors in both layers.

\begin{figure}[H]
    \centering
    \includegraphics[scale=0.5]{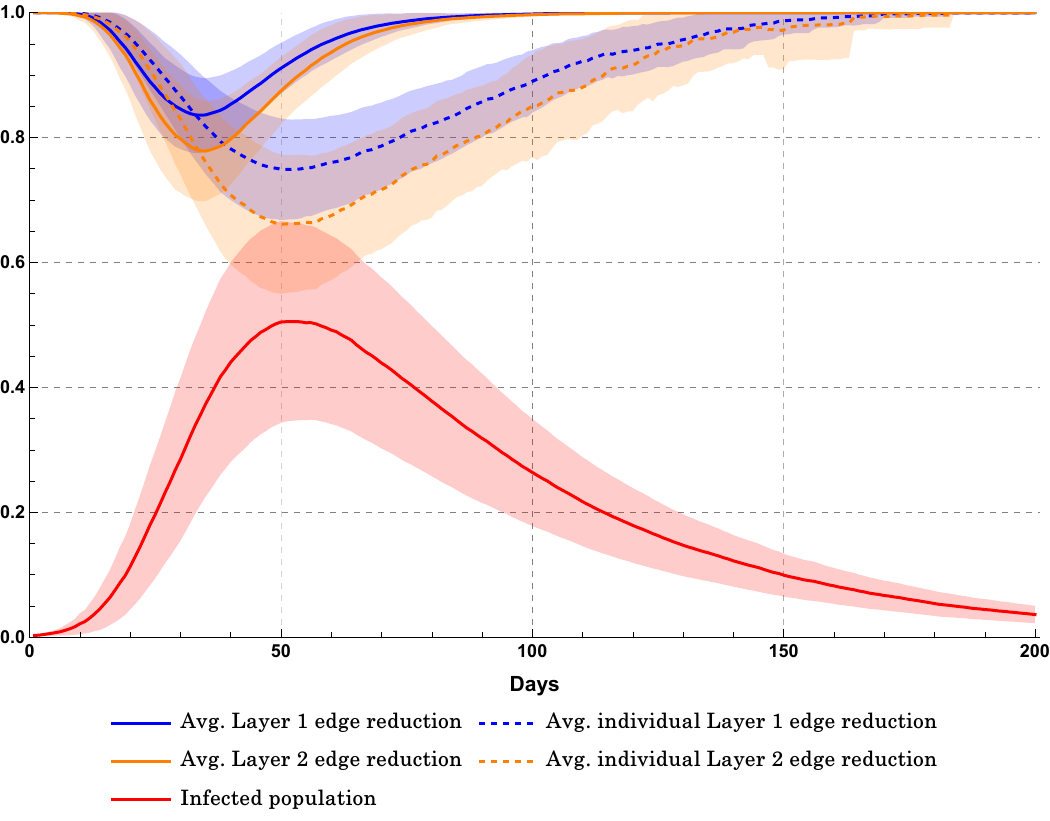}
    \caption{Contact reduction and infected population for an adaptive network model with two interposed layers, each as a random model network. The overall disease peak coincides with the maximum effort susceptibles perform with respect to each layer.}
    \label{fig:TwoLayersPlot}
\end{figure}

Figure \ref{fig:TwoLayersPlot} displays the contact adaptation that occurs for each network. This simulation involves two decision processes, one for each layer (network). In this example, the first network is a Erd\"{o}s–R\'{e}nyi model with $n=500$ nodes and probability of connection of $p_1=0.05$. In contrast, the second layer is a similar model with $p_2=0.02$, resulting in a lower network connectivity. We use adaptive parameters of $\nu_1 = 0.025, \nu_2 = 0.01, T_1 = T_2 = 14$ days, and disease parameters of $\beta=0.009$ and $\gamma= 0.02$. We can see that the susceptible effort consists of an adaptation of contacts with both networks, and the maximum effort for susceptibles occurs asynchronously with respect to the maximum global effort, but synchronously with respect to the disease behavior. Adaptation from the second layer is more severe, as the adaptive parameters allow for a higher sensitivity towards contacts in this layer. This example works as a parallel implementation of the main adaptive framework. Where individuals make two contact reduction decisions, each informed by different risk perceptions. In this example, the general susceptible individual is more receptive to diminishing contacts with the outer social network than with the immediate or closer layer.

\subsection{Deciding sporadic contacts} 
\begin{figure}[H]
    \centering
    \begin{tikzpicture}[scale=1, 
    suscnode/.style={person, scale=1},
    basenode/.style={person, shirt=black, scale=1},
    infnode/.style={person, shirt=red, skin=green, scale=1},]
    
    \filldraw[blue!10] (3.25,5.75) circle (65pt);

    \node[basenode] (A) at (3,6) {};
    \node[suscnode] (B) at (2,7) {};
    \node[suscnode] (C) at (5,5) {};
    \node[infnode] (D) at (3.5,4) {};
    \node[suscnode] (E) at (2,5.2) {};

    \node[suscnode] (G) at (5.551,2) {};
    \node[suscnode] (H) at (7,3) {};
    \node[infnode] (I) at (8,5.75) {};
    \node[suscnode] (J) at (2.25,2.5) {};
    
    \draw[->, color=black!70] (A) -- (B);
    \draw[->, color=black!70] (A) -- (C);
    \draw[->, color=black!70] (A) -- (D);
    \draw[->, color=black!70] (A) -- (E);

    \draw[->, color=black, style=dashed] (A) -- (G);
    \draw[->, color=black, style=dashed] (A) -- (I);
    \draw[-, color=gray, style=dashed] (I) -- (H);
    \draw[-, color=gray, style=dashed] (C) -- (I);
    \draw[-, color=gray, style=dashed] (D) -- (J);
    \draw[-, color=gray, style=dashed] (E) -- (J);
    \draw[-, color=gray, style=dashed] (H) -- (J);
    \draw[-, color=gray, style=dashed] (H) -- (G);

    \node[text width=2.5cm] at (9,4) 
    {Outer sporadic contacts};

    \node[text width=1.5cm] at (3.8,7.3) 
    {Neighbor contacts};

\end{tikzpicture}
\caption{Two types of contacts within a single network. Neighbor contacts serve as the first layer of regular contacts. Other nodes outside serve as a layer of sporadic contacts.}
\label{fig:sporadic_contacts_model}
\end{figure}
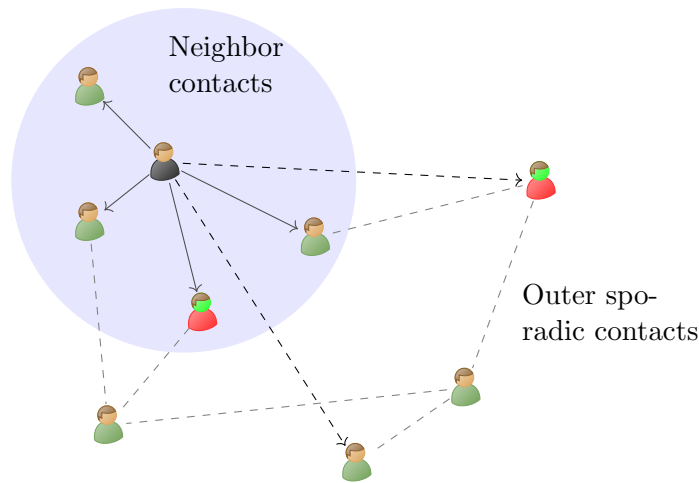

Another possibility of modelling irregular contacts is to use an underlying network and make susceptible individuals engage with random, non-immediate, neighbors. In \cite{Sanchez23}, the selection of these sporadic contacts is made by taking into account the individual's geographical county and then having them engage with random contacts from neighboring counties, using an uniform distribution based on each county's population. Here we offer a simpler example, in which the sporadic contacts are taken from within the base network, as a sample of non-neighbor nodes. Like in the previous example, the individual is then faced with two decision processes: compute the optimal number of neighbors to engage with, based on the local information, and decide the number of sporadic individuals to contact, based on the network global disease information, see Figure \ref{fig:sporadic_contacts_model} for an illustration of both decision layers. The first process, of deciding contacts within the core group, is performed with local information: the adaptive decision uses the individual's utility function, their neighbors, and the infected nodes in this neighborhood. For the sporadic decision process, although it also performs a decision process using the adaptive framework presented in this article, the decision is made here based on global information on infected nodes in the entire network. Consequently, this example assumes that all nodes are equally accessible as contacts, and their epidemic status is equally obtainable as well.

\begin{figure}[ht!]
  \centering
  \begin{subfigure}[b]{0.48\linewidth}
    \centering
    \includegraphics[width=\linewidth]{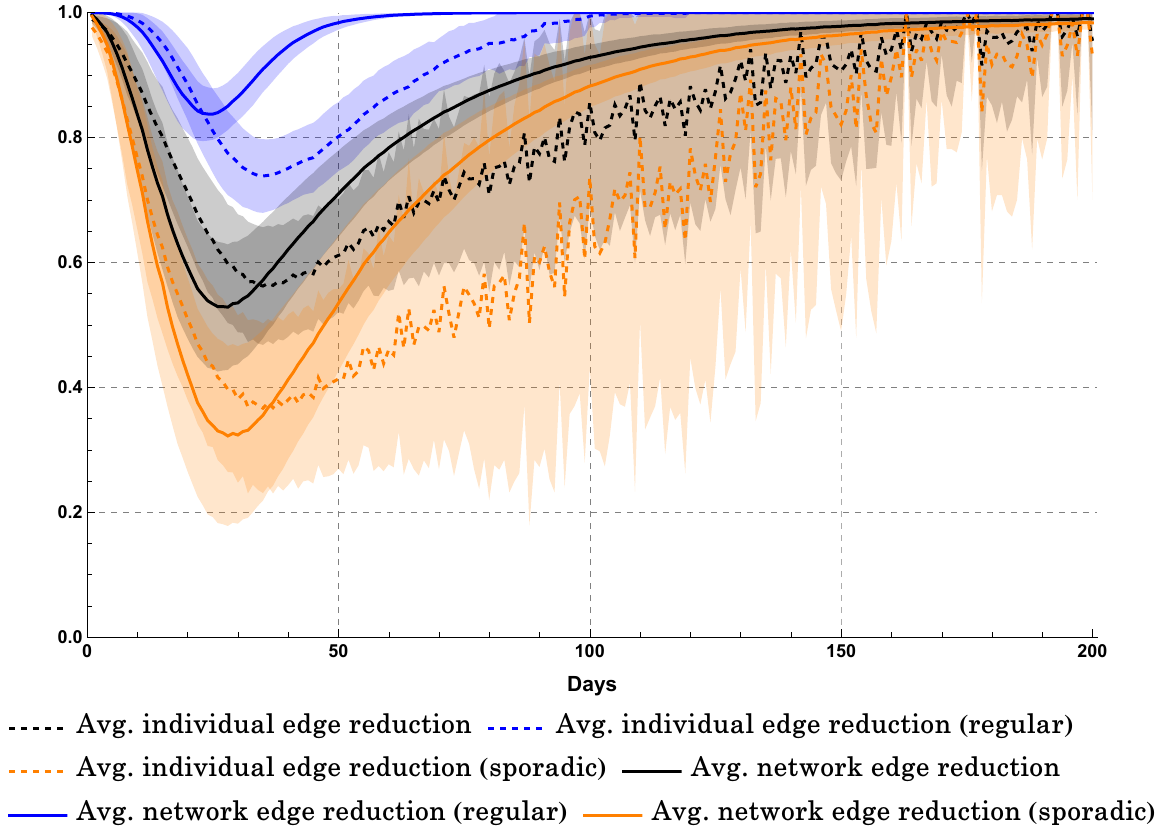}
    \caption{Global and local behavioral responses.}
    \label{fig:sporadic_network_a}
  \end{subfigure}
  \hfill
  \begin{subfigure}[b]{0.48\linewidth}
    \centering
    \includegraphics[width=\linewidth]{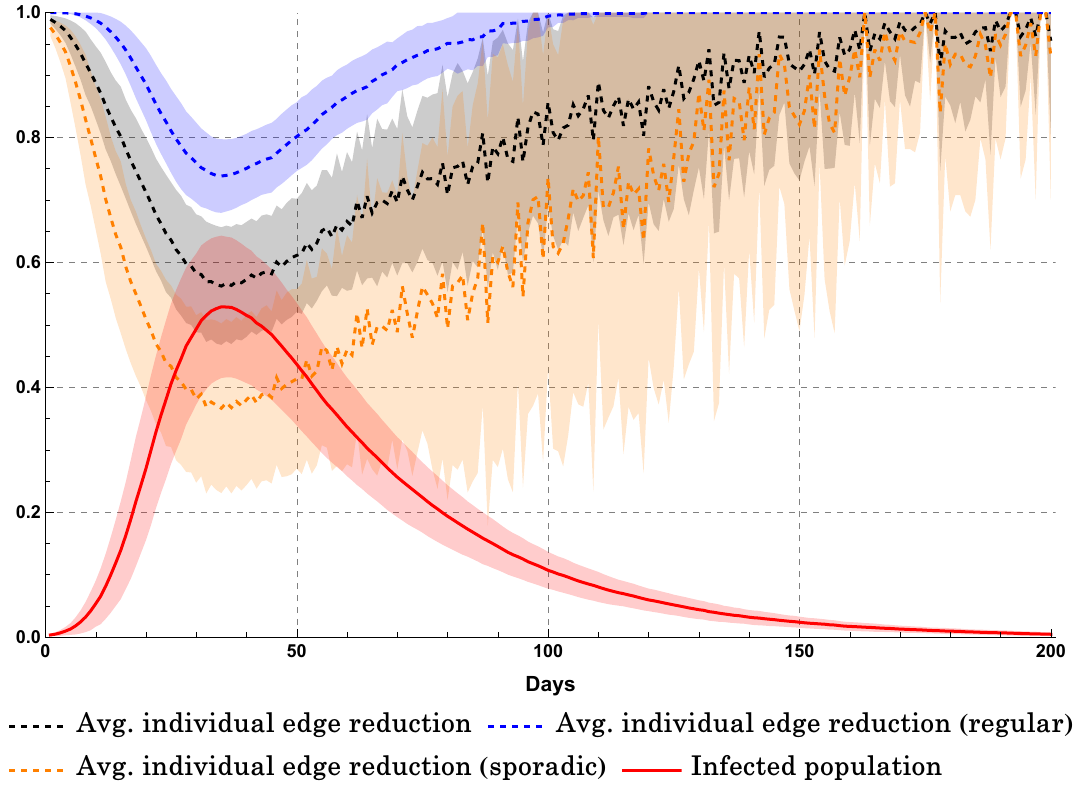}
    \caption{Population dynamics and local behavioral response.}
    \label{fig:sporadic_network_b}
  \end{subfigure}
  \caption{Population dynamics and global/local contact adaptation for an adaptive network model with two contact types: regular contacts (from network neighbors) and sporadic contacts (from non-neighbor nodes). Two adaptation processes emerge, and the global response averages both. Maximum effort aligns with the disease peak for both contact types.}
  \label{fig:sporadic_network}
\end{figure}

Figure \ref{fig:sporadic_network} shows the global/local behavioral response for this layer model, as well as the resulting disease dynamics. This simulation was performed using an underlying random model network of $500$ nodes and a node connectivity probability of $p=0.025$. The neighbor layer decision process was performed with adaptive parameters $T=7, \nu_1=0.05$ and the sporadic layer used $T=7,\nu_2=0.025$, disease parameters were $\beta=0.02, \gamma=0.03$. 

The higher risk perception for sporadic contacts results in a higher edge reduction with respect to sporadic contacts than with neighboring contacts. In this case, we observe a more irregular behavior for susceptible individuals when it comes to deciding their sporadic contacts as the disease progresses. In this example, each day a susceptible node contacts several random nodes outside its immediate neighborhood. The number of contacts is determined by the global prevalence of infection. The random selection might result in a node selecting all healthy random contacts one day, and encountering an infected node in this random selection the next day, which brings about this irregularity in the average response as the number of susceptible nodes becomes lower. However, in general, the sporadic contact response returns to an original maximum as the infected population decreases.

Note that the maximum reduction effort done at the individual level is again synchronous with the disease prevalence peak, for both regular and sporadic contact types. This reinforces the main result of this article in other, more complex applications of network modeling, such as the use of contact layers or sporadic edge connections.


\section*{Network parameters analysis}


\begin{table}[H]
\small
    \centering
    \begin{tabular}{|l|l|}
        \hline
        \textbf{Network Model} & Erd\H{o}s–R\'{e}nyi $(p=0.05)$ \\
        \hline
        Size & 500 \\
        Average Clustering Coefficient & 0.0501 \\
        Average Degree Centrality & 0.0501 \\
        Average Closeness Centrality & 0.4508 \\
        Node Connectivity & 11.55 \\
        \hline
        \textbf{Network Model} & Albert–Barabási $(m=20)$ \\
        \hline
        Size & 500 \\
        Average Clustering Coefficient & 0.1515 \\
        Average Degree Centrality & 0.0770 \\
        Average Closeness Centrality & 0.5023 \\
        Node Connectivity & 19.5 \\
        \hline
        \textbf{Network Model} & Watts–Strogatz $(k=15, p=0.05)$ \\
        \hline
        Size & 500 \\
        Average Clustering Coefficient & 0.6000 \\
        Average Degree Centrality & 0.0281 \\
        Average Closeness Centrality & 0.2661 \\
        Node Connectivity & 11.2 \\
        \hline
    \end{tabular}
    \caption{Average network metrics at time $t=0$ for the graph models used in this study. Each value is the average over 100 simulations.}
    \label{tab:networkmetricstab}
\end{table}

\begin{figure}[H]
    \centering
    \includegraphics[scale=0.5]{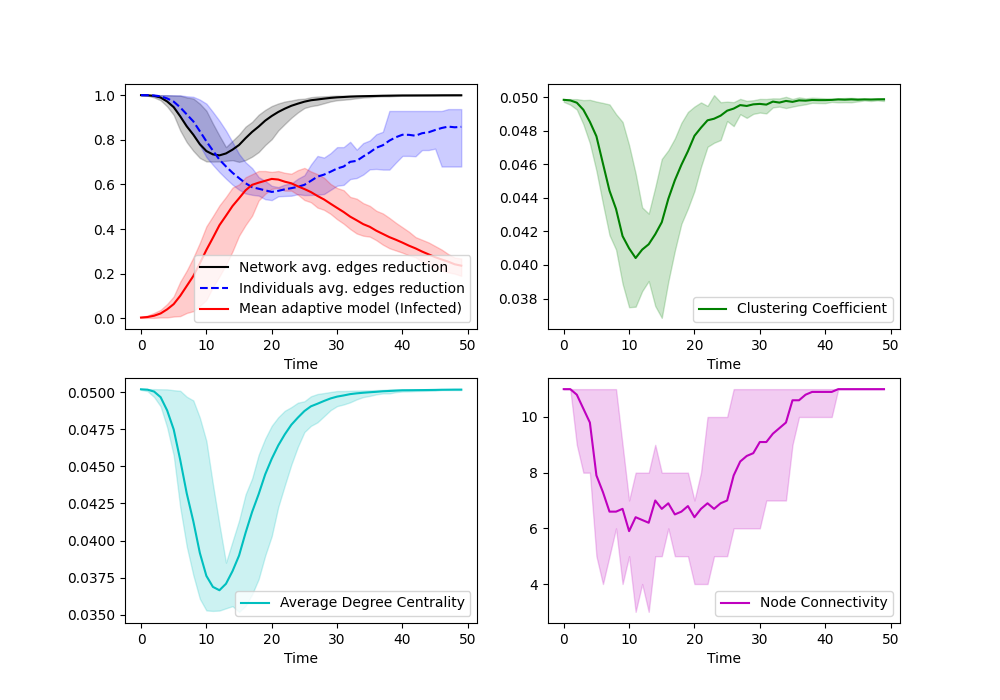}
    \caption{Metrics of underlying network during the epidemic development. We see that clustering, connectivity, and centrality coefficients decrease when edge reduction occurs at the network level.}
    \label{fig:networkmetricsfig}
\end{figure}


\end{document}